\documentclass[fleqn,usenatbib]{mnras}

\usepackage{newtxtext,newtxmath}

\usepackage[T1]{fontenc}

\DeclareRobustCommand{\VAN}[3]{#2}
\let\VANthebibliography\thebibliography
\def\thebibliography{\DeclareRobustCommand{\VAN}[3]{##3}\VANthebibliography}

\usepackage{graphicx}	
\usepackage{amsmath}	
\usepackage{booktabs}
\usepackage{pdflscape}
\usepackage{rotating}
\usepackage{xcolor}

\title[DUVET: Outflow Scaling Relations]{DUVET: sub-kiloparsec resolved star formation driven outflows in a sample of local starbursting disk galaxies}

\author[B. Reichardt Chu et al.]{
Bronwyn Reichardt Chu,$^{1,2,3}$\thanks{E-mail: bronwyn.j.reichardtchu AT durham.ac.uk}
Deanne B. Fisher,$^{1,2}$ 
John Chisholm,$^{4}$ 
Danielle Berg,$^{4}$
Alberto Bolatto,$^{5}$
\newauthor{
Alex J. Cameron,$^{6}$
Drummond B. Fielding,$^{7,8}$
Rodrigo Herrera-Camus,$^{9}$ 
Glenn G. Kacprzak,$^{1,2}$
Miao Li,$^{10}$
}
\newauthor{
Anna F. McLeod,$^{3,11}$
Daniel K. McPherson,$^{1,2}$
Nikole M. Nielsen,$^{12,1,2}$
Ryan J. Rickards Vaught,$^{13}$
}
\newauthor{
Sophia G. Ridolfo,$^{14,1,2}$
and Karin Sandstrom$^{15}$
}
\\
$^{1}$Centre for Astrophysics and Supercomputing, Swinburne University of Technology, Hawthorn, VIC 3122, Australia\\
$^{2}$ARC Centre of Excellence for All Sky Astrophysics in 3 Dimensions (ASTRO 3D), Australia\\
$^{3}$Centre for Extragalactic Astronomy, Department of Physics, Durham University, South Road, Durham DH1 3LE, UK\\
$^{4}$Department of Astronomy, University of Texas, Austin, TX 78712, USA\\
$^{5}$University of Maryland, College Park, MD 20742, USA\\
$^{6}$Sub-department of Astrophysics, University of Oxford, Keble Road, Oxford, OX1 3RH, UK\\
$^{7}$Center for Computational Astrophysics, Flatiron Institute, 162 Fifth Avenue, New York, NY 10010, USA\\
$^{8}$Department of Astronomy, Cornell University, Ithaca, NY 14853, USA\\
$^{9}$Departamento de Astronom\'ia, Universidad de Concepci\'on, Barrio Universitario, Concepci\'on 4070032, Chile\\
$^{10}$Institute for Astronomy, School of Physics, Zhejiang University, 866 Yuhangtang Road, Hangzhou, 310027, China\\
$^{11}$Institute for Computational Cosmology, Department of Physics, Durham University, South Road, Durham DH1 3LE, UK\\
$^{12}$Homer L. Dodge Department of Physics and Astronomy, The University of Oklahoma, 440 W. Brooks St., Norman, OK 73019, USA\\
$^{13}$Space Telescope Science Institute, 3700 San Martin Drive, Baltimore, MD 21218, USA\\
$^{14}$Center for Astrophysics, Harvard \& Smithsonian, 60 Garden Street, Cambridge, MA 02138, USA\\
$^{15}$Department of Astronomy \& Astrophysics, University of California, San Diego, 9500 Gilman Drive, La Jolla, CA, 92093, USA\\
}

\date{Accepted XXX. Received YYY; in original form ZZZ}

\pubyear{2024}

\begin{document}
\label{firstpage}
\pagerange{\pageref{firstpage}--\pageref{lastpage}}
\maketitle

\begin{abstract}
We measure resolved (kiloparsec-scale)  outflow properties in a sample of 10 starburst galaxies from the DUVET (Deep near-UV observations of Entrained gas in Turbulent galaxies) sample, using Keck/KCWI observations of H$\beta$ and [OIII]~$\lambda$5007. We measure $\sim460$ lines-of-sight that contain outflows, and use these to study scaling relationships of outflow velocity ($v_{\rm out}$), mass-loading factor ($\eta$; mass outflow rate per SFR) and mass flux ($\dot{\Sigma}_{\rm out}$; mass outflow rate per area) with co-located SFR surface density ($\Sigma_{\rm SFR}$) and stellar mass surface density ($\Sigma_{\ast}$). We find strong, positive correlations of $\dot{\Sigma}_{\rm out} \propto \Sigma_{\rm SFR}^{1.2}$ and $\dot{\Sigma}_{\rm out} \propto \Sigma_{\ast}^{1.5}$. We also find shallow correlations between $v_{\rm out}$ and both $\Sigma_{\rm SFR}$ and $\Sigma_{\ast}$. Our resolved observations do not suggest a threshold in outflows with $\Sigma_{\rm SFR}$, but rather we find that the local specific SFR ($\Sigma_{\rm SFR}/\Sigma_\ast$) is a better predictor of where outflows are detected. We find that outflows are very common above $\Sigma_{\rm SFR}/\Sigma_\ast\gtrsim 0.1$~Gyr$^{-1}$ and rare below this value. We argue that our results are consistent with a picture in which outflows are driven by supernovae, and require more significant injected energy in higher mass surface density environments to overcome local gravity. The correlations we present here provide a statistically robust, direct comparison for simulations and higher redshift results from JWST. 
\end{abstract}




\section{Introduction}

\defcitealias{walcher2009stellarpopulation}{W09}
\defcitealias{bruzual2003bc03stellarpopulation}{BC03}
\defcitealias{reichardtchu2022resolvedmaps}{RC22a}
\defcitealias{reichardtchu2022spatiallyresolved}{RC22b}

The evolution of galaxies is shaped by the baryon cycle. Cold gas is accreted onto galaxies, used as fuel in star formation which in turn enriches the gas, and is then ejected from the galaxy to enrich the surrounding environment \citep{somerville2015physicalmodels}.  Star formation-driven outflows are a necessary component of this cycle, contributing to the enrichment of the circumgalactic medium (CGM) \citep{tumlinson2017CGMreview, cameron2021duvetMrk1486}, and suppressing star formation through the removal of gas \citep{veilleux2005galacticwinds, bolatto2013suppression_ngc253, reichardtchu2022spatiallyresolved}.  Galaxy-wide outflows are required for simulations to reproduce basic galaxy properties including the galaxy mass function, typical galaxy sizes, and the Kennicutt-Schmidt Law \citep[e.g][]{springel2003cosmologicalsim, oppenheimer2006cosmologicalsims, hopkins2012stellarfeedback, hopkins2014FIRE}.  To constrain the implementation of galaxy-wide outflows in simulations, the simulations need to be compared to empirical measurements of outflow quantities.  It is, therefore, necessary that we understand the observational properties of outflows and their driving mechanisms.

Star formation-driven outflows have been observed across cosmic time \citep[e.g.][]{heckman2000absorptionline, chen2010absorption, rubin2010persistence, davies2019kiloparsec}.  Outflows are an observational tracer of the feedback process that regulates star formation, preventing runaway star formation in multiple ways.  First, it is expected that the gravitational weight of the disk creates pressure in the interstellar medium (ISM) which is balanced by the energy and momentum injected into the ISM by young massive stars and supernovae.  This injected energy and momentum creates turbulence, suppressing the star formation occurring within the galaxy disk \citep[e.g.][]{ostriker2010regulation, fauchergiguere2013feedbackregulatedsf, hayward2017stellar, krumholz2018unifiedmodel, ostriker2022pressureregulated}.  Recent observations that compare velocity dispersion to the star formation properties find correlations that are consistent with feedback-based regulation of star formation in disks \citep{fisher2019testing,girard2021systematic}. Pre-supernova feedback from young massive stars affects the density of the ISM gas that the supernovae then explode into, regulating the impact of the supernova feedback \citep{mcleod2021impactpreSNefeedback, chevance2022preSNefeedback}. In more extreme starburst environments, clustered supernovae create expanding ``superbubbles'' which drive galactic winds when they break out of the disk \citep{fielding2018clustered, kim2018solarneighborhood,vijayan2020kinematics, orr2022burstingbubbles_letter}.
Once these expanding gas bubbles break out of the disk, the gas they drive out of the galaxy is no longer available for star formation.  Scaling relations are useful for identifying where this occurs, and in which local regions the outflow dominates over turbulence as the main method of star formation regulation.  Moreover, establishing observational scaling relations provides direct tests for simulations \citep[e.g.][]{kim2020firstresultssmaug, rathjen2023silccVIIgaskinematics}.

Simulations provide outflow properties such as the outflow velocity $v_{\rm out}$, the outflow mass flux $\Dot{\Sigma}_{\rm out}$, and the mass loading factor $\eta$ as a function of the star formation rate surface density $\Sigma_{\rm SFR}$  that are heavily dependent on their underlying assumptions of how stellar feedback is implemented in their (often subgrid) models \citep[e.g.][]{nelson2019firstresultsTNG50, kim2020firstresultssmaug, pandya2021characterisingFIRE2}.  
Characterising these scaling relationships provides constraints on our understanding of the physical drivers of outflows.  
For example, the relationship between the star formation rate surface density and the outflow velocity provides constraints on the primary driving mechanism of the outflows.  Outflows primarily driven by mechanical energy from supernovae are expected to have a shallow relationship as $v_{\rm out} \propto \Sigma_{\rm SFR}^{0.1}$ \citep{chen2010absorption, li2017supernovaedriven, kim2020firstresultssmaug}. 
However, if the outflows are primarily driven by momentum given to the gas by radiation from young massive stars, then the expected relationship has a steeper dependence as $v_{\rm out}\propto\Sigma_{\rm SFR}^2$ \citep{murray2005momentumwinds, kornei2012properties, hopkins2012stellarfeedback}.  Outflows are almost certainly driven by a combination of both of these mechanisms. Spatial and temporal differences in the preprocessing of the ISM material that SNe explode into and entrain will contribute to the observed $v_{\rm out}-\Sigma_{\rm SFR}$ relationship. 
In addition, supernovae generate cosmic rays which are also expected to play a role in driving outflows \citep[e.g.][]{girichidis2016cosmicrayoutflows, girichidis2024spectrallyresolvedCRs}.  The extent to which this is important is unclear due to uncertainties in the parameters of cosmic ray transport \citep[e.g.][]{naab2017theoreticalchallenges, crocker2021cosmicrays_outflows, kim2023introducingTIGRESS-NCR}.
To drive realistic galaxy-wide outflows, simulations need to include empirically motivated prescriptions that include the effects from all of the relevant feedback processes.  As an initial constraint, these very different power-laws make observational tests of the primary contributor of the two models possible.


Observations of outflows have historically been limited by their low surface brightness to studies of either integrated galaxy samples, or stacked galaxy samples. 
Trends between the outflow velocity and global galaxy measurements of stellar mass, SFR and $\Sigma_{\rm SFR}$ of galaxies have been reported by many studies \citep[e.g.][]{martin2005mapping, rupke2005outflowsdiscussion, steidel2010structure, chen2010absorption, newman2012sins, kornei2012properties, arribas2014ionisedgasoutflows, bordoloi2014dependence, Rubin2014evidence, chisholm2015scaling, chisholm2017mass, forsterschreiber2019kmos3d}.
However, galaxy-wide observations have an underlying dependence on the stellar mass of the galaxy \citep{newman2012sins, chisholm2015scaling, nelson2019firstresultsTNG50}, making it difficult to test the predicted local correlations. Moreover, it has not been established how the resolved measurements may alter empirical correlations between outflow and the galaxy from which they launched. For example, if outflows are launched from small-scale regions in a galaxy, then global averages may wash out any existing local relationships. Alternatively, it may be that aggregate effects of the wind dominate the outflow energetics, and thus global averages are more important. 

Resolved observations of face-on galaxies allow the comparison of outflow properties to co-located galaxy properties.  Using resolved observations, we can trace the outflowing kinematics back to the local energy and momentum injected by star formation to test on which scales the star formation drives galaxy-scale outflows.
Recently developed sensitive IFUs, such as KCWI, MUSE and JWST/NIRSpec, have made this possible. 
In \citet{reichardtchu2022resolvedmaps} (hereafter RC22a) we studied resolved ionised outflows in the pilot target for the DUVET sample, IRAS~08339+6517.  We found a shallow correlation between the local co-located $\Sigma_{\rm SFR}$ and $v_{\rm out}$, consistent with trends expected from supernovae being the primary outflow driving mechanism.  In our follow-up paper, \citet{reichardtchu2022spatiallyresolved} (hereafter RC22b), we found a relationship between the star formation rate surface density and outflow mass flux of $\Dot{\Sigma}_{\rm out}\propto\Sigma_{\rm SFR}^{1.06\pm0.1}$.  In this paper, we extend this analysis to a sample of ten face-on galaxies from the DUVET survey.

The paper is organised as follows.  We describe our galaxy sample in Section~\ref{subsec:duvet}, and our observations and data reduction of the sample in Sections~\ref{subsec:obs} and \ref{subsec:cont_subtract}.  We describe our method to fit for outflows in Section~\ref{subsec:threadcount}. The resulting relationships of the maximum outflow velocity, mass outflow flux, and mass loading factor with SFR surface density and stellar mass surface density are explored in Section~\ref{sec:results}.  Conclusions are presented in Section~\ref{sec:conclusions}.  Throughout the paper, we assume a flat $\Lambda$CDM cosmology with $H_0=69.3$~km~Mpc$^{-1}$~s$^{-1}$ and $\Omega_0=0.3$ \citep{hinshaw2013wmap9}.

\section{Observations and Data Reduction}
\label{sec:data_obs}

\subsection{DUVET sample}
\label{subsec:duvet}

DUVET (Deep near-UV observations of Entrained gas in Turbulent galaxies) is a sample of starbursting disk galaxies at $z\sim0.02-0.04$. 
The sample contains both face-on and edge-on galaxies so that we can build a three-dimensional understanding of the impact of stellar feedback and star formation-driven outflows on galaxies with high star formation rate surface densities, $\Sigma_{\rm SFR}$, and their environments
(e.g. face-on: \citetalias{reichardtchu2022resolvedmaps, reichardtchu2022spatiallyresolved}; edge-on: \citealt{cameron2021duvetMrk1486, mcpherson2023duvet_mrk1486}, McPherson et al. in prep). The galaxies are chosen to have total SFRs at minimum $5\times$ the main sequence value for their total stellar mass.  The galaxies are also required to have the morphology and kinematics of a disk so that we can remove the underlying velocity field.  The total sample has 27 galaxies and a stellar mass range of $10^9-10^{11}$~M$_\odot$.

This work uses 10 targets from the DUVET sample that have low inclinations. Inclination is estimated photometrically using the ratio of the major-to-minor axis in near-IR broadband images from 2MASS H-band (1.7~$\mu$m) and Spitzer/IRAC Ch1 (3.6~$\mu$m). Nine of these targets have inclinations of $i<15^\circ$, and one target is moderately inclined (CGCG~453-062: $i=54^\circ$).  The inclined target was observed due to a range in LST at the time of observation that did not have sufficient targets available that meet the DUVET sample selection criterion.  We will highlight CGCG~453-062 in our discussion of results and in the main results plots.

The ten galaxies chosen for this work span four orders-of-magnitude in $\Sigma_{\rm SFR}$ ($\sim0.001-10$~M$_\odot$~yr$^{-1}$~kpc$^{-2}$) in $\sim500$~pc scale regions, while keeping the total galaxy stellar mass $10-20\times10^{10}$~M$_\odot$. 
Properties for all 10 galaxies examined here are given in Table~\ref{tab:gal_params}, in order of increasing total stellar mass.  We have calculated the infrared SFR (Column 5) using Band 4 ($22\mu$m) photometry from the WISE Telescope \citep{wright2010WISE} given in the AllWISE Source Catalog \citep{cutri2014allWISEdatarelease}, following the relationship from \citet[][their Eq. 7]{cluver2017calibratingsf}.  We have used the magnitudes given in column \texttt{w4gmag} of the Catalog, which cross-matches the WISE source positions with the 2MASS Extended Source Catalog \citep{skrutskie20062MASS} and fits an elliptical aperture based on the 2MASS data, scaled for the larger WISE Band-4 point spread function. This allows for the extended, non-circular profiles of our galaxies. 


\begin{table*}
    \centering
    \begin{tabular}{lcccccccccc}
        \hline 
        Name & $z$ & M$_*$ & SFR$_{\rm IFU, H\beta}^*$ & SFR$_{\rm total, IR}$ & $A_v$ & $A_\lambda$ & $r_{50}$ & $r_{90}$ & $\log$~[OIII]/H$\beta$ & References \\ 
        & & ($10^{10}~M_\odot$) & ($M_\odot$~yr$^{-1}$) & ($M_\odot$~yr$^{-1}$) & (mag) & (mag) & ($\arcsec$) & ($\arcsec$) & & \\
        (1) & (2) & (3) & (4) & (5) & (6) & (7) & (8) & (9) & (10) & (11)\\
        \hline
        IRAS~08339+6517 & 0.019 & 1.10 & $7.59\pm0.05$ & $12.8\pm0.1$ & $0.27^{+0.32}_{-0.27}$ & 0.365 & 2.6 & 5.90 & $0.12\pm0.03$
            & 1\\
        IRAS~15229+0511 & 0.036 & 3.02  & $4.65\pm0.05$ & $14.4\pm0.1$ & $0.57^{+1.01}_{-0.57}$ & 0.183 & 3.7 & 9.10 & $-0.35\pm0.01$
            & 2\\
        KISSR~1084 & 0.032 & 3.05  & $2.62\pm0.02^*$ & $3.69\pm0.06$ & $0.75\pm0.59$ & 0.188 & 4.8 & 8.20 & $-0.35\pm0.02$
            & 3\\
        NGC~7316 & 0.019 & 4.17  & $2.23\pm0.01^*$ & $3.25\pm0.04$ & $0.49\pm0.40$ & 0.186 & 6.8 & 12.3 & $-0.22\pm0.02$
            & 4\\
        UGC~01385 & 0.019 & 4.74  & $5.68\pm0.06$ & $12.8\pm0.1$ & $0.32^{+0.67}_{-0.32}$ & 0.316 & 2.8 & 7.60 & $-0.37\pm0.03$
            & 5\\
        UGC~10099 & 0.035 & 5.73 & $4.84\pm0.03$ & $8.17\pm0.06$ & $0.37\pm0.32$ & 0.069 & 2.5 & 6.20 & $-0.04\pm0.03$
            & 6 \\
        CGCG~453-062 & 0.025 & 8.99  & $15.3\pm0.1^*$ & $10.9\pm0.08$ & $0.10\pm0.80$ & 0.395 & 7.2 & 13.2 & $-0.47\pm0.01$
            & 5 \\
        UGC~12150 & 0.021 & 11.0  & $3.43\pm0.03^*$ & $14.7\pm0.08$ & $0.86^{+1.20}_{-0.86}$ & 0.271 & 5.3 & 11.1 & $-0.35\pm0.01$
            & 5 \\
        IRAS~20351+2521 & 0.034 & 11.2  & $16.3\pm0.1^*$ & $28.9\pm0.2$ & $0.88\pm0.65$ & 0.751 & 7.3 & 12.1 & $-0.18\pm0.02$
            & 7\\
        NGC~0695 & 0.032 & 20.1 & $33.4\pm0.3^*$ & $29.9\pm0.2$ & $1.66\pm0.72$ & 0.351 & 5.9 & 11.2 & $-0.26\pm0.01$
            & 5 \\
        \hline 
    \end{tabular}
    \caption[]{
    Columns are: (1) Galaxy name.  
    (2) Redshift.  
    (3) Stellar mass from literature, references given in Column~10.  
    (4) Star formation rate from ionised gas using the H\ensuremath{\beta} line.  
    (5) Star formation rate from WISE Band 4 data, following \protect\citet[][their Eq.7]{cluver2017calibratingsf}. 
    (6) Average extinction in the disk gas based on the H\ensuremath{\gamma}/H\ensuremath{\beta} flux ratio within $r_{90}$. 
    (7) Galactic extinction in the band WFC3~F390W from \citet{schlafly2011measuringreddening}.
    (8) and (9) effective radius and 90\% radius in arcseconds from g-band PanSTARRS data respectively. 
    (10) Median \ensuremath{\log_{10}(}[OIII]~\ensuremath{\lambda5007/}H\ensuremath{\beta)} flux ratio.
    (11) References: 
    1--\cite{fisher2022extremevariation}, 
    2--\cite{bik2022spatiallyresolved},
    3--\cite{cook2019census}
    4--\cite{fernandezlorenzo2013stellarmasssize}, 
    5--\cite{howell2010All-skyLIRGsurvey}, 
    6--\cite{kouroumpatzakis2021starformationreferencesurvey}, 
    7--\cite{shangguan2019interstellarmedium}.
    \\
    $^*$ We use the total flux measured in H\ensuremath{\beta} to calculate SFR\ensuremath{_{\rm IFU, H\beta}}, however, some galaxies in our sample extend beyond the FOV of our observations and so this is a lower limit of the total SFR for the galaxy. The affected SFR values are indicated using an asterisk.
    }
    \label{tab:gal_params}
\end{table*}

\subsection{Observations}
\label{subsec:obs}

The galaxies were observed with KCWI/Keck II \citep{Morrissey2018kcwi} over a range of nights given in Table~\ref{tab:gal_obs}, with sub-arcsecond seeing conditions.  We used the BM grating ($R\sim2500$) in the large IFU slicer mode, giving a field of view of $20\farcs4\times33\farcs$ with spatial sampling of $0\farcs29\times1\farcs35$, which is seeing limited in the short spaxel side direction.  The galaxies were observed using a half-slice dither pattern in the direction of the long side of the spaxel to allow for better spatial sampling. The exposure time and number of exposures for each galaxy are given in Column~3 of Table~\ref{tab:gal_obs}. All galaxies are observed a ``blue" and ``red" grating setting. In this paper we only use the red setting. The central wavelength set for the BM grating of the red setting is given in Column~4 of Table~\ref{tab:gal_obs} for each galaxy and was chosen to include all wavelengths from H$\gamma$ to [OIII]~$\lambda5007$. The galaxies were also observed with a ``blue'' configuration, to extend the wavelength coverage from H$\gamma$ to below the [OII] doublet.  The observations using the ``blue" configuration are described in McPherson et al. {\em in prep}.  
A separate sky field was observed beyond the virial radius for each galaxy either directly before or after the science exposures.  The observations and data reduction for the pilot target, IRAS~08339+6517 were discussed in detail in \citetalias{reichardtchu2022resolvedmaps} and \cite{fisher2022extremevariation}.  
The observations and data reduction for the remaining nine galaxies follow a similar method. 

The data were reduced using the standard IDL KCWI Data Extraction and Reduction Pipeline (Version 1.1.0)\footnote{\url{https://github.com/Keck-DataReductionPipelines/KcwiDRP}}.  Before combining the images, small-scale imperfections in the WCS were accounted for by a re-alignment based on the H$\gamma$ emission line, which is detected in all images and allows for alignment with bluer wavelength settings taken on each galaxy not used in this work. 
The H$\gamma$ flux was compared in each pixel using an iterative minimisation method, and the WCS of the images was adjusted to result in the minimum average residual across each galaxy.  
The python package \textsc{Montage}\footnote{\url{http://montage.ipac.caltech.edu/}} was then used to reproject the images to produce square spaxels of $0\farcs29\times0\farcs29$, based on the shorter side length of the original rectangular spaxels.  
The resulting reprojected images of each galaxy were co-added in \textsc{Montage} using the adjusted WCS coordinates.  The variance cubes were similarly reprojected and coadded.  The number of images and their exposure times are given in Column~3 of Table~\ref{tab:gal_obs} for each galaxy.  The reduction process for DUVET galaxies is described in more detail in \citet{mcpherson2023duvet_mrk1486}.

To account for the (on average) $0\farcs7$ seeing conditions, the resulting realigned, reprojected and coadded cubes were binned $3\times3$ to obtain cubes with a final spaxel size of $0\farcs87\times0\farcs87$, which is comparable to the typical seeing FWHM.  The dither pattern of the observations in the direction of the long side of the original spaxels ($0\farcs29\times1\farcs35$) allows for the reconstructed images to better sample the seeing.
\citetalias{reichardtchu2022resolvedmaps} found stronger correlations between outflow and co-located galaxy parameters above a re-binned spaxel size of ${\sim0.5}$~kpc, and discussed possible geometric and temporal causes (see \citetalias{reichardtchu2022resolvedmaps} for a discussion). 
The reprojected spaxel size for the galaxies in this work is equivalent to $0.34$~kpc~$\times~0.34$~kpc at $z=0.019$ for our closest galaxies and $0.63$~kpc~$\times~0.63$~kpc at $z=0.036$ for our furthest galaxy.  
Our pilot target IRAS~08339+6517 is the exception to this, retaining its original data reduction from \citetalias{reichardtchu2022resolvedmaps} resulting in rectangular spaxels of $0\farcs3\times1\farcs35$ ($0.1$~kpc$\times0.5$~kpc at $z=0.019$).

\begin{table*}
    \centering
    \begin{tabular}{lccccc}
        \hline 
        Name & RA & Dec & Obs date & Exp times & Central wavelength \\
         & (J2000) & (J2000) & & (s) & (\AA) \\
        (1) & (2) & (3) & (4) & (5) & (6) \\
        \hline 
        IRAS~08339+6517 & 08:38:23.18s & +65:07:15.2s & 15 Feb 2018 & 1200, 600, 300, 100 & 4800\\
         CGCG~453-062 & 23:04:56.53s & +19:33:08.0s & 20 Oct 2019 & $6\times300$ & 4800 \\
         UGC~01385 & 01:54:53.79s & +36:55:04.6s & 20 Oct 2019 & $6\times300$ & 4800 \\
         UGC~12150 & 22:41:12.26s & +34:14:57.0s & 21 Oct 2019 & $6\times300$ & 4800 \\
         NGC~0695 & 01:51:14.24s & +22:34:56.5s & 21 Oct 2019 & $4\times300$ & 4800 \\
         NGC~7316 & 22:35:56.34s & +20:19:20.1s & 20 \& 21 Oct 2019 & $1\times200, 6\times300, 3\times500$ & 4800 \\
         IRAS~15229+0511 & 15:25:27.49s & +05:00:29.9s & 22 March 2020 & $6\times300$ & 4850 \\
         UGC~10099 & 15:56:36.40s & +41:52:50.5s & 22 March 2020 & $6\times300$ & 4850 \\
         IRAS~20351+2521 & 20:37:17.72s & +25:31:37.7s & 16 May 2020 & $6\times300$ & 4820 \\
         KISSR~1084 & 16:49:05.27s & +29:45:31.6s & 16 May 2020 & $6\times300$ & 4820 \\
         \hline 
    \end{tabular}
    \caption{Columns are:~(1) Galaxy name. 
    (2) and (3) RA and Dec from NED*.
    (4) The date the galaxy was observed with KCWI/Keck II. 
    (5) Number and length of exposure times in seconds.  
    (6) is the set central wavelength of the BM grating. \\
    *The NASA/IPAC Extragalactic Database (NED) is funded by the National Aeronautics and Space Administration and operated by the California Institute of Technology.
    }
    \label{tab:gal_obs}
\end{table*}

\section{Methods}
\label{sec:methods}

\subsection{Continuum Subtraction}
\label{subsec:cont_subtract}

Before each cube was continuum subtracted, foreground extinction due to the Milky Way was corrected in each spaxel using the \citet{cardelli1989relationship} law and galactic extinction values from the  \citet{schlafly2011measuringreddening} recalibration of the \citet{schlegel1998mapsofdust_SFD} extinction map.  The value used for each galaxy is given in Table~\ref{tab:gal_params}.

The full spectrum fitting code \texttt{pPXF} \citep{cappellari2017ppxf} was used to fit the stellar continuum.  
For this fitting to occur, the continuum was required to have a signal-to-noise greater than 3 in a continuum band (4600~\AA~$<\lambda<4800$~\AA) for each galaxy data cube.  For nine of the ten galaxies in this work, the continuum fitting used semi-empirical templates from \citet{walcher2009stellarpopulation}.  IRAS~08339+6517 was found to have a lower gas-phase metallicity than the rest of the sample based on a comparison of its emission line ratios \citep[e.g.][]{lopezsanchez2006IRAS08paper}.  Due to this difference, and to keep consistency with the results from \citetalias{reichardtchu2022resolvedmaps} and \citetalias{reichardtchu2022spatiallyresolved}, the continuum subtraction for IRAS~08339+6517 used the BPASS templates \citep[Version 2.2.1,][]{stanway2018reevaluatingbpass}, including binary systems with a broken power-law initial mass function with a slope of -1.3 between $0.1~M_\odot$ and $1.0~M_\odot$, a slope of -2.35 above $1.0~M_\odot$, and an upper limit of $300~M_\odot$.   
The BPASS templates were built to cover younger ages and lower metallicities than the \citet{walcher2009stellarpopulation} templates.   

Post continuum subtraction, we identified some spaxels where a wide residual absorption feature near H$\beta$ is not fully removed through the continuum subtraction process. This does not occur in all spaxels, and appears to be more common in more heavily extincted systems.  The continuum subtraction of starburst galaxies is unreliable, due to well-known issues in creating stellar population libraries to model the continuum in these environments \citep{conroy2013SED_review}.  We therefore address any regions that were poorly fit by fitting any remaining residuals.  Using short wavelength bands on either side of the emission lines, we fit a linear function across [OIII]~$\lambda5007$ and a quadratic across H$\beta$ such that the corrected spectrum has a flat baseline.  The wavelengths of the bands on either side of the emission lines are carefully chosen for each galaxy to exclude any broad emission from the emission line, and the correction is interpolated across the baseline.  We use a quadratic function across H$\beta$ as it is more likely to be affected by residual hydrogen absorption in stellar photospheres.  The resulting baselines are visually inspected for a subset of spaxels in each galaxy.  Examples for  H$\beta$ and [OIII]~$\lambda$5007 are given in Appendix~\ref{appendix:baseline_corr}. Including the baseline increased the total number of spaxels containing evidence of outflows found across all 10 galaxies from 458 to 465, resulting in an increase in the median covering fraction within $r_{50}$ of only $\sim$0.01~dex. We tested our results with and without this additional baseline correction and found the change fell within the error bars for the difference in both median outflow velocity and median outflow mass flux.  
Including the baseline correction decreased the median mass loading factor by $\sim2\%$.


Finally, an internal extinction correction was calculated on a spaxel-by-spaxel basis for all galaxies using the emission line ratio H$\beta$/H$\gamma$ and the \cite{calzetti2001dustopacity} extinction curve. Across our sample, we have a median signal-to-noise ratio of 29 in H$\beta$ and 8.5 in H$\gamma$. This extinction correction was applied when calculating the star formation rate and the mass outflow rate.  The impacts of including an extinction correction within the outflow mass calculation are discussed further in Section~\ref{subsec:mass_flux}.


\subsection{Emission-line Fitting Method}
\label{subsec:threadcount}

\begin{figure*}
    \centering
    \includegraphics[width=0.9\textwidth]{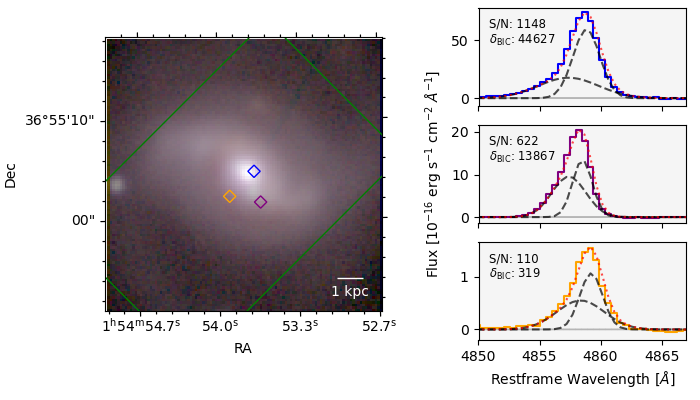}
\caption{Example fits from DUVET target UGC~01385.  Left panel: an RGB image of UGC~01385 (R: Pan-STARRS z-band, G: Pan-STARRS r-band, B: Pan-STARRS g-band).  Large green rectangle shows the footprint of our KCWI observations.  Blue, purple and orange small squares show highlighted KCWI spaxels, for which the H$\beta$ emission line fits are shown in the right-hand panels.  Right-hand panels: for each panel, data is in blue, purple or orange corresponding to the squares in the left panel.  The variance is shown as a grey band. Dashed black lines show the two-component Gaussian fits.  Dotted red line shows the complete fitted model.  The signal-to-noise and $\delta_{\rm BIC}$ are given for each spaxel (see Sec.~\ref{subsec:threadcount} for more discussion of the $\delta_{\rm BIC}$).}
    \label{fig:ex_fits}
\end{figure*}

To identify outflows we follow the method developed in previous DUVET papers \citep{reichardtchu2022resolvedmaps,mcpherson2023duvet_mrk1486}.  We implement multi-component Gaussian fitting using the Python package \texttt{threadcount}\footnote{\url{https://github.com/astrodee/threadcount}}, which uses the Python fitting package \texttt{lmfit} \citep{newville2019lmfit0.9.14} with a Nelder-Mead simplex minimisation algorithm.  We fit both a single and a double Gaussian to the H$\beta$ and [OIII]~$\lambda$5007 emission lines independently in spaxels where we measure a signal-to-noise greater than 10 in the emission line.  To determine whether the extra Gaussian component is necessary, we use the Bayesian information criterion (BIC). The form of the BIC we use here is the Aikake information criterion, the special case where each model is fit to the same number of data points, defined as
\begin{equation}
    BIC = \chi^2 + 2 N_{\rm variables},
    \label{eq:BIC}
\end{equation}
where $\chi^2=\sum(f_{\rm data}-f_{\rm model})^2/\sigma^2$, and 
$\sigma$ is the data uncertainty.  $N_{\rm variables}$ is the number of variables included in the fit, which is 4 for single Gaussian fits and 7 for double Gaussian fits, where for each fit three variables describe each Gaussian component and the extra variable is for an additional linear component.  Some example fits are shown for one of our targets, UGC~01385, in Fig.~\ref{fig:ex_fits}.   

The double Gaussian model is only chosen where it gives a lower BIC value than the single Gaussian model minus a threshold value such that $BIC_{\rm double Gaussian}<BIC_{\rm single Gaussian} - \delta_{\rm BIC}$.  
The typical literature value used for the BIC threshold is $\delta_{\rm BIC}=-10$ \citep{Kass1995BayesFactors, swinbank2019energetics, avery2021incidence}.  However, this threshold is based on the assumption that the model errors are independent and distributed according to a normal distribution.  Emission lines in typical spiral galaxies have been shown to be better fit by multiple Gaussian components \citep{ho2014sami}.  Observations of singular H~{\footnotesize II} regions also find emission lines require multiple Gaussian components \citep{rozas2007halphalineprofiles} or a Voigt profile \citep{keto2008earlyevolution, galvanmadrid2012ALMAandVLAobs}.
Our data has extremely high signal-to-noise.  We select only spaxels that have a median signal-to-noise per wavelength channel of $\gtrsim10$, implying a peak signal-to-noise of 100-200 across many of the emission lines.  This is sufficiently high to identify any fluctuations away from a Gaussian shape in the galaxy emission, even with the limited spectral resolution.  
In addition, the models available for continuum subtraction are imperfect representations of young, starburst populations, and residuals from this process may be mistaken for a low-flux broad component if a lenient $\delta_{\rm BIC}$ is used.  In \citetalias{reichardtchu2022resolvedmaps} we found that using the typically adopted threshold value of $\delta_{\rm BIC}$ results in all spaxels requiring multiple Gaussian component models, even those spaxels in which a by-eye examination favours only a single Gaussian component.  We, therefore, choose a stricter value for $\delta_{\rm BIC}$ using the following method.

\begin{figure*}
    \centering
    \includegraphics[width=0.8\textwidth]{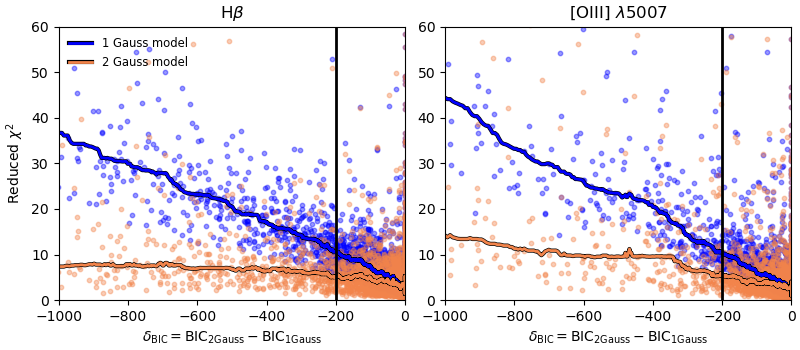}
    \caption{The reduced $\chi^2$ value plotted against the difference in BIC for fits to H$\beta$ (left) and [OIII]~5007 (right) emission lines.  Blue points represent fits using a simple 1 Gaussian model.  Orange points represent fits to the same data using a 2 Gaussian model.  Solid lines of the same colour show the rolling median of the points.  The vertical black line is the threshold value used for the whole sample, $\delta_{\rm BIC}=-200$.  At this threshold, the two populations of fits are statistically different.}
    \label{fig:delta_BIC}
\end{figure*}

To find the optimum threshold value to use with our data, we first ran \textsc{threadcount} with a lenient threshold of $\delta_{\rm BIC}=-10$ for all galaxies in our sample.  We then compared the reduced $\chi^2$ ($\chi^2_{red}=\chi^2/\nu$, where $\nu$ is the degrees of freedom) value from fits to the emission lines using both a simple 1 Gaussian model and a 2 Gaussian model to the $\delta_{\rm BIC}$ returned comparing the two models.  There is some degree of circularity in this reasoning for any individual emission line fit, as the $\chi^2$ value is calculated as part of the BIC.  However, we are using the comparative difference in $\chi^2_{red}$ value in order to understand the BIC values.
\ref{fig:delta_BIC} shows the $\chi^2_{red}$ value plotted against the $\delta_{\rm BIC}$ for all galaxies, with fits to H$\beta$ represented in the left, and [OIII]~$\lambda5007$ in the right panels respectively.

For each galaxy, we determined the threshold to be the maximum $\delta_{\rm BIC}$ where the running median of the $\chi^2_{red}$ for the 1 Gaussian models becomes one standard deviation away from the running median of the $\chi^2_{red}$ for the 2 Gaussian models.
We then took the average across all galaxies, finding an average H$\beta$ threshold of -161, and an average [OIII]~$\lambda5007$ threshold of -132.  
We make a conservative choice to round to $\delta_{\rm BIC}=-200$ for our threshold value across the sample, and this is plotted as a black vertical line in Fig.~\ref{fig:delta_BIC}.
This conservative choice may bias results by removing fits to low signal-to-noise spectra, as well as fits suggesting low-velocity winds.

We restrict the double Gaussian fits such that the bluer component must have a smaller amplitude and larger velocity dispersion than the redder component.  When a double Gaussian model is the statistically preferred fit, we assume the broad, blue-shifted component measures flux from outflowing gas and the narrower, higher amplitude component measures flux from gas within the galaxy disk (see \citetalias{reichardtchu2022resolvedmaps, reichardtchu2022spatiallyresolved} and \citealt{mcpherson2023duvet_mrk1486} for more detail). A visual inspection of the data did not find any redshifted secondary peaks. The software prompts the user to confirm fits with marginal BIC values or those in which the central velocity shift of the second component is small. Likewise, we visually inspect a large fraction of resulting fits, and generally find good fits. 


To calculate the uncertainties on the fits, \texttt{threadcount} runs a Monte Carlo simulation.  The spectral pixels of each input emission line are adjusted using a normal distribution with a standard deviation from the observed variance in our data.  The resulting simulated emission line is then re-fit, and the uncertainty in the fit for that spaxel is calculated as the standard deviation of all simulated spectra fits.  In this paper, we perform 100 simulations of each emission line (H$\beta$ and [OIII]~$\lambda5007$) for each spaxel.

\begin{figure*}
    \centering
    \includegraphics[width=\linewidth]{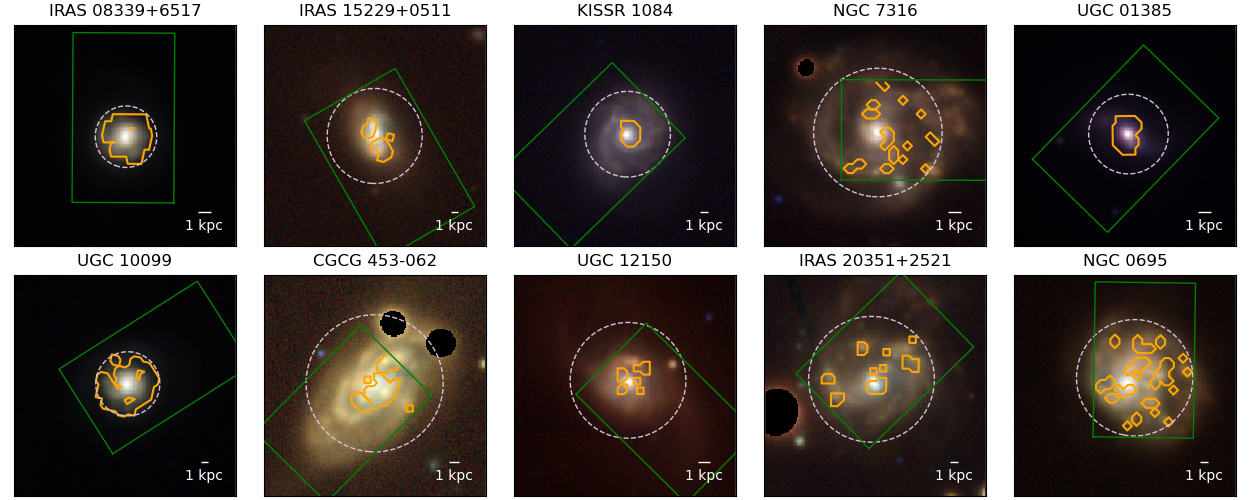}
    \caption{Maps of the location of the outflows for all 10 galaxies in our sample.  RGB images are from Pan-STARRS (R: z-band, G: r-band, B:g-band), using $42\farcs5 \times42\farcs5$ cutouts. Galaxies are organised by increasing stellar mass, from left to right, top line before bottom line (galaxy masses given in Table~\ref{tab:gal_params}). Large green rectangle shows the footprint of our KCWI observations ($20\farcs4\times33\farcs$).  Light purple dashed circle shows $r_{90}$ (given for each galaxy in Table~\ref{tab:gal_params}).  Orange contours show where outflows have been detected in each galaxy. A 1~kpc scale bar is in white in the bottom right of each panel. Bright foreground objects have been masked for NGC~7316, CGCG~453-062 and IRAS~20351+2521.}
    \label{fig:outflow_maps}
\end{figure*}

We restrict the spaxels in our results to those within $r_{90}$ of each galaxy (given in Table~\ref{tab:gal_params}) to exclude any spaxels on the edge of the galaxy that may have more complicated kinematics due to inflowing gas. 
We restrict the velocity dispersion of both components to be greater than the instrument dispersion of KCWI, $\sigma_{\rm inst}=41.9$~km~s$^{-1}$, such that all outflows have resolved dispersions, and the outflow velocity dispersion is greater than the galaxy dispersion. We additionally exclude 
12 spaxels in the centre of CGCG~453-062 which are double-peaked, making decomposing the outflow component unreliable.  
Maps of the final results for the location of spaxels with evidence of outflows for each galaxy are given in Fig.~\ref{fig:outflow_maps}.

\section{Results: Star formation driven outflow scaling relations}
\label{sec:results}

In this section, we compare results from fitting for outflows in each spaxel to the co-located galaxy properties.  We derive scaling relationships for the maximum outflow velocity, the outflow mass flux and the mass loading factor with star formation rate surface density and stellar mass surface density.  A summary of these correlations is given in Table~\ref{tab:correlations}.  We also discuss the relationship of the fraction of spaxels containing outflows with total galaxy stellar mass and global star formation rate surface density, and compare our summed galaxy results to total galaxy measurements from the literature.  

\begin{table*}
    \centering
    \begin{tabular}{c|c|c|c|c|c|c|c}
        \hline 
         $x$ & $y$ & $\alpha$ & $\beta$ & $R_P$ & $p$-value & $R_S$ & $p$-value \\
         \hline 
         $\Sigma_{\rm SFR}$ & $v_{\rm out, H\beta}$ & 2.48$\pm$0.01 & 0.20$\pm$0.01 & 0.44 & $2\times10^{-23}$ & 0.48 & $2\times10^{-28}$\\
         $\Sigma_{\rm SFR}$ & $v_{\rm out, [OIII]}$ & 2.52$\pm$0.01 & 0.12$\pm$0.01 & 0.27 & $4\times10^{-7}$ & 0.37 & $4\times10^{-12}$ \\
         
         $\Sigma_{\rm SFR}$ & $\dot{\Sigma}_{\rm out, extcorr}$ & 0.53$\pm$0.03 & 1.21$\pm$0.03 & 0.84 & $8\times10^{-128}$ & 0.82 & $2\times10^{-117}$\\
         $\Sigma_{\rm SFR}$ & $\dot{\Sigma}_{\rm out, nocorr}$ & 0.10$\pm$0.04 & 1.08$\pm$0.04 & 0.75 & $1\times10^{-86}$ & 0.71 & $1\times10^{-74}$ \\
         
         $\Sigma_{\rm SFR}$ & $\eta$ & 0.03$\pm$0.04 & -0.03$\pm$0.04 & -0.04 & $0.33$ & -0.02 & $0.72$ \\


         $\Sigma_\ast$ & $v_{\rm out, H\beta}$ & 1.66$\pm$0.05 & 0.22$\pm$0.02 & 0.49 & $4\times10^{-31}$ & 0.53 & $1\times10^{-37}$ \\
         $\Sigma_\ast$ & $v_{\rm out, [OIII]}$ & 2.15$\pm$0.07 & 0.09$\pm$0.02 & 0.19 & $4\times10^{-4}$ & 0.28 & $2\times10^{-7}$ \\
         
         $\Sigma_\ast$ & $\dot{\Sigma}_{\rm out, nocorr}$ & $-4.62\pm$0.14 & 1.30$\pm$0.04 & 0.79 & $3\times10^{-101}$ & 0.83 & $3\times10^{-120}$\\
         $\Sigma_\ast$ & $\eta$ & $-1.44\pm$0.13 & 0.49$\pm$0.04 & 0.44 & $2\times10^-23$ & 0.51 & $5\times10^{-33}$\\
         \hline 
    \end{tabular}
    \caption{Power law fits to the given parameters for spatially resolved data from all 10 galaxies in our sample, of the form $\log_{10}(y)=\alpha + \beta\log_{10}(x)$, using the method of orthogonal distance regression.  $R_P$ is the Pearson correlation coefficient, with its $p$-value in the next column to the right.  $R_S$ is the Spearman's Rank correlation coefficient for the given correlation, with its $p$-value in the next column to the right. The subscripts `extcorr' and `nocorr' indicate where the outflow has been extinction corrected the same as the disk gas, or not extinction corrected, respectively.}
    \label{tab:correlations}
\end{table*}

\subsection{SFR surface density and maximum outflow velocity}
\label{subsec:outflow_velocity}

The kinematics of the outflow and their relationship to the SFR surface density can be used to distinguish between subgrid physical models describing the primary launching mechanism of the outflow. \citetalias{reichardtchu2022resolvedmaps} found a shallow relationship between the SFR surface density and the maximum outflow velocity, $v_{\rm out}\propto\Sigma_{\rm SFR}^{0.1-0.2}$, consistent with outflows driven by the energy from supernovae for IRAS~08339+6517.  In this paper, we extend this analysis to nine more face-on galaxies from the DUVET sample. 

We measure the star formation rate, SFR, in each spaxel using the narrow line flux from fits to the extinction corrected H$\beta$ emission line.  In the optical, star formation rates are typically inferred from H$\alpha$ luminosities \citep[e.g.][]{kennicutt2012star}.  Given that our observations do not include the H$\alpha$ emission line, we scale our observations of H$\beta$ by the lab value for the $F_{\rm H\alpha}/F_{\rm H\beta}$ flux ratio such that
\begin{equation}
    {\rm SFR} = C_{\rm H\alpha}\left(\frac{F_{\rm H\alpha}}{F_{\rm H\beta}}\right) 10^{0.4A_{\rm H\beta}} F_{\rm H\beta},
    \label{eq:sfr}
\end{equation}
where $C_{\rm H\alpha}=5.5335\times10^{-42}~M_\odot~$yr$^{-1}~($erg~s$^{-1})^{-1}$ is the scale parameter assuming a \citet{kroupa2003imf} initial mass function \citep[IMF;][]{hao2011dustcorrected}.  $F_{\rm H\alpha}/F_{\rm H\beta}=2.87$ is the flux ratio \citep[$T_e=10^4$~K and Case B recombination,][]{osterbrock2006astrophysics}.  $A_{\rm H\beta}$ is the extinction derived from the observed H$\beta$/H$\gamma$ ratio and a \citet{calzetti2001dustopacity} attenuation curve (see Section~\ref{subsec:cont_subtract}).  $F_{\rm H\beta}$ is the observed H$\beta$ flux, using flux from the narrow component of the fits. This assumes that emission from the broad component is due to outflowing gas, and is not caused by star formation. \citetalias{reichardtchu2022resolvedmaps} found that including flux associated with outflowing gas causes an average increase in the measured $\Sigma_{\rm SFR}$ of $\sim25\%$.

We report the total SFR for each galaxy from the H$\beta$ emission line in Column~4 of Table~\ref{tab:gal_params}.  We note that for six of the galaxies in our sample (CGCG~453-062, UGC~12150, NGC~0695, NGC~7316, IRAS~20351+2521 and KISSR~1084) the galaxy extends beyond the FOV of our KCWI pointing.  The SFR that we measure is therefore a lower limit for the total SFR of these galaxies.

Star formation rate surface density, $\Sigma_{\rm SFR}$, is calculated using the spaxel size for each measurement.  
We find that the average spaxel within $r_{90}$ across all galaxies in the sample has a median $\log(\Sigma_{\rm SFR} [$M$_\odot$~yr$^{-1}$~kpc$^{-2}])$ of -1.3 with a 1$\sigma$ span of $\pm0.9$~dex.    
Spaxels across all galaxies where we find evidence for outflows in both [OIII]~$\lambda5007$ and H$\beta$ have a median $\log(\Sigma_{\rm SFR} [$M$_\odot$~yr$^{-1}$~kpc$^{-2}])$ of -0.9 with a 1$\sigma$ span of $\pm0.8$~dex.  
The typical sub-kpc region hosting outflows has, therefore, roughly half an order-of-magnitude higher $\Sigma_{\rm SFR}$ than the median across all 10 galaxies.

Following \citetalias{reichardtchu2022resolvedmaps}, we define the maximum outflow velocity as 
\begin{equation}
    v_{\rm out} = |v_{\rm narrow} - v_{\rm broad}| + 2\sigma_{\rm broad},
    \label{eq:max_outflow_vel}
\end{equation}
where $v_{\rm narrow}$ and $v_{\rm broad}$ are the fitted Gaussian centres in velocity space for narrow and broad Gaussians respectively. $\sigma_{\rm broad}$ is the standard deviation of the broad Gaussian. The instrumental velocity dispersion of 0.7\AA\ (41.9~km~s$^{-1}$) is removed in quadrature. 
The median $v_{\rm out}$ error is of order $\sim30$~km~s$^{-1}$ and $\sim40$~km~s$^{-1}$ for H$\beta$ and [OIII]~$\lambda5007$ respectively.
We note that by construction of how we fit the data, the outflow velocity should implicitly be assumed to have a negative sign, as it is the blue-shifted outflow component of the emission lines.

All of the galaxies in our sample except for IRAS~08339+6517 are brighter in H$\beta$ than in [OIII]~$\lambda5007$, with the median $\log_{10}($[OIII]~$\lambda5007/$H$\beta)$ ratio for the galaxy emission (excluding outflows) given in Table~\ref{tab:gal_params}.


We note that CGCG~453-062 has a high inclination angle ($i\approx54^\circ$), and therefore the measured outflow velocities require correction.  We deproject the velocities by dividing by $\cos{i}$.  
For more detail on the velocity deprojection, see Appendix~\ref{appendix:ind_gals}.
We use the deprojected velocity in the remainder of this work for CGCG~453-062.  The difference in velocities after correction for galaxies with inclinations $<15^\circ$ is of order $\sim5$~km~s$^{-1}$, which is negligible. We, therefore, do not deproject the velocities of the rest of the galaxies as their inclination angles are below $15^\circ$.  

\begin{figure*}
    \centering
    \includegraphics[width=\textwidth]{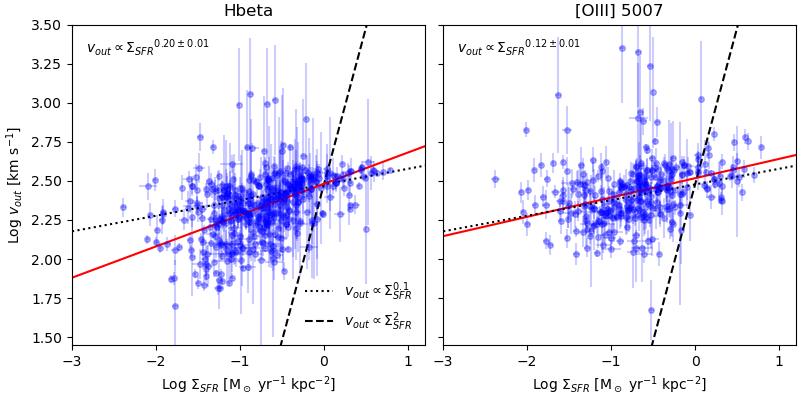}
    \caption{The maximum outflow velocity, $v_{\rm out}$, plotted against the star formation rate surface density, $\Sigma_{\rm SFR}$, for spaxels of sub-kpc resolution in 10 local starbursting disk galaxies which required a double Gaussian fit for H$\beta$ (447 spaxels, left panel) and [OIII]~$\lambda5007$ (324 spaxels, right panel) emission lines.  For relations for individual galaxies, see Appendix~\ref{appendix:ind_gals}. 
    The fit to the data is given in the top left corner and shown as a solid red line in each panel.  
    The dashed line represents a model where outflows are primarily driven by momentum injected from young massive stars \citep[$v_{\rm out}\propto\Sigma_{\rm SFR}^2$;][]{murray2011radiationpressure}.  The dotted line represents a model where outflows are primarily driven by energy injected from supernovae \citep[$v_{\rm out}\propto\Sigma_{\rm SFR}^{0.1}$;][]{chen2010absorption}.  For observations of both H$\beta$ and [OIII]~$\lambda5007$, the fitted correlation is not consistent with the momentum-driven model.}
    \label{fig:sigsfr_vout_allgals}
\end{figure*}


In Fig.~\ref{fig:sigsfr_vout_allgals} we show the results from our full sample for both the H$\beta$ and [OIII]~$\lambda5007$ emission lines, which were fit independently.   For figures for individual galaxies, see Appendix~\ref{appendix:ind_gals} (Figures \ref{fig:sigsfr_vout_individual_hbeta} and \ref{fig:sigsfr_vout_individual_OIII}).  In both H$\beta$ and [OIII]~$\lambda5007$ we measure the outflow velocity $v_{\rm out}$ across almost three orders of magnitude in $\Sigma_{\rm SFR}$.
We find median outflow velocities of $229$~km~s$^{-1}$ with a 1$\sigma$ scatter of 120~km~s$^{-1}$, and $274$~km~s$^{-1}$ with a 1$\sigma$ scatter of 218~km~s$^{-1}$ for the H$\beta$ and [OIII]~$\lambda5007$ velocities respectively.  We find a Pearson correlation coefficient of $R_P=0.43$ ($p-$value$\ll10^{-3}$) 
for the relationship between $\log(\Sigma_{\rm SFR})$ and $\log(v_{\rm out})$ for H$\beta$, and $R_P=0.27$ ($p-$value$\ll10^{-3}$) for [OIII]~$\lambda5007$.  

Using the method of orthogonal distance regression\footnote{Scipy's ODR fitter, see \url{https://docs.scipy.org/doc/scipy/reference/odr.html}} we fit a relationship between $\Sigma_{\rm SFR}$ and $v_{\rm out}$ for the results from H$\beta$ and [OIII]~$\lambda5007$ separately, and find 
\begin{equation}
    v_{\rm out, H\beta}=10^{2.48\pm0.01}\Sigma_{\rm SFR}^{0.20\pm0.01}
    \label{eq:sigma_sfr_vout_hbeta}
\end{equation}
for velocities found from fitting the H$\beta$ emission lines, and
\begin{equation}
    v_{\rm out, OIII}=10^{2.52\pm0.01}\Sigma_{\rm SFR}^{0.12\pm0.01}
    \label{eq:sigma_sfr_vout_OIII}
\end{equation}
for velocities found from fitting the [OIII]~$\lambda5007$ emission lines. These fits are shown in Fig.~\ref{fig:sigsfr_vout_allgals} as solid red lines.
We compare these to two models commonly used for the relationship between $v_{\rm out}$ and $\Sigma_{\rm SFR}$ in simulations.  The dashed line shows the expected correlation if the outflows are primarily driven by the radiation pressure giving momentum to the gas surrounding young massive stars \citep[$v_{\rm out}\propto\Sigma_{\rm SFR}^2$;][]{murray2011radiationpressure}.  The dotted line shows the expected correlation if the outflows are primarily driven by the energy injected from supernovae \citep[$v_{\rm out}\propto\Sigma_{\rm SFR}^{0.1}$;][]{chen2010absorption}.  
We recover shallow relationships for both [OIII]~$\lambda5007$ and H$\beta$ maximum outflow velocities, which are not consistent with the model for a purely ``momentum-driven'' outflow.

\citet{avery2021incidence} fitted for ionised gas outflows using kinematically tied fits to H$\beta$, H$\alpha$, and the [OIII]~$\lambda\lambda4959, 5007$, [NII]~$\lambda\lambda6548, 6583$, and [SII]~$\lambda\lambda6716, 6731$ doublets in radially binned star-forming galaxies with disk morphologies from the MaNGA Survey.  Excluding galaxies with AGN-driven outflows, they found a shallow negative relationship of $v_{\rm out}\propto\Sigma_{\rm SFR}^{-0.07\pm0.02}$.  On the other hand, using fits to the H$\alpha$ and [NII]~$\lambda\lambda6548, 6583$ emission lines, \citet{davies2019kiloparsec} found a steeper relationship of $v_{\rm out}\propto\Sigma_{\rm SFR}^{0.34\pm0.10}$ for $z\sim2.3$ galaxies from the SINS/zC-SINF AO Survey stacked by $\Sigma_{\rm SFR}$.
Our results lie between these two studies.
We, however, have allowed H$\beta$ and [OIII]~$\lambda5007$ to be fit with kinematically independent fits. 
We note that the MaNGA galaxies cover a wide range of $\Sigma_{\rm SFR}$, but are not chosen to be starbursting as our sample of galaxies is. The sample of SINS galaxies in \citet{davies2019kiloparsec} has a median offset of 2$\times$ the main sequence SFR at $z=2-2.5$, and covers a similar range of $\Sigma_{\rm SFR}$ as the sample of galaxies studied here.

\citet{zheng2024ionisedoutflows_LMC} made resolved observations of the Large Magellanic Cloud (LMC) measure outflowing ionised gas in absorption using stars from the HST ULLYSES dataset. 
Combining their resolved data with galaxy-integrated observations of outflows in local starburst galaxies from the CLASSY sample \citep{xu2022CLASSYproperties}, \citet{zheng2024ionisedoutflows_LMC} found a shallow relationship of $v_{\rm out}\propto\Sigma_{\rm SFR}^{0.23\pm0.03}$. This is consistent with our results, and is in close agreement with our $v_{\rm out}-\Sigma_{\rm SFR}$ relationship for the outflows measured with H$\beta$.

The galaxies in our sample do not have the same signal-to-noise in H$\beta$ as in [OIII]~$\lambda5007$ and are on average brighter in H$\beta$.  The higher S/N enables us to fit the outflow component more reliably in H$\beta$.  The outflow component is unresolved in [OIII]~$\lambda5007$ for the galaxies IRAS~15229+0511, NGC~7316, and UGC~12150.
Note that excluding spaxels where we observe an outflow in only one emission line restricts the results to spaxels with ${\Sigma_{\rm SFR}>10^{-2}~M_\odot}$~yr$^{-1}$~kpc$^{-2}$, only 0.3~dex higher than the minimum $\Sigma_{\rm SFR}$ when all outflow detections are included.  
The fits for $v_{\rm out}-\Sigma_{\rm SFR}$ are consistent regardless of whether we restrict our data.  We do find that the Pearson correlation coefficient for the outflow velocities measured in H$\beta$ increases from $R_P=0.43$ to $R_P=0.47$ ($p-$value$\ll10^{-3}$), 
and for the [OIII]~$\lambda5007$ results from $R_P=0.27$ to $R_P=0.32$ ($p-$value$\ll10^{-3}$).  

For those spaxels with both H$\beta$ and [OIII] broad components, the outflow velocities measured for [OIII]~$\lambda5007$ are on average $\sim20$~km~s$^{-1}$ higher than those measured for H$\beta$.  The velocity is measured to be 256~km~s$^{-1}$ with a 1$\sigma$ scatter of 104~km~s$^{-1}$ in H$\beta$ and 275~km~s$^{-1}$ with a 1$\sigma$ scatter of 214~km~s$^{-1}$ in [OIII]~$\lambda5007$.   This velocity difference is more apparent at lower $\Sigma_{\rm SFR}$.  
This difference could be due to systematics introduced by the absorption line around H$\beta$.  Alternatively, it is plausible that [OIII]~$\lambda5007$ is associated with a higher ionisation state than H$\beta$, and this gas is intrinsically moving faster.

\subsection{SFR surface density and outflow mass flux}
\label{subsec:mass_flux}

It is expected that regions of galaxies with a higher SFR surface density will drive a higher mass outflow rate, $\Dot{M}_{\rm out}$, which has been observed in resolved observations of one target \citepalias{reichardtchu2022spatiallyresolved}.  
The mass outflow rate is defined as
\begin{equation}
    \Dot{M}_{\rm out} = \frac{1.36m_{\rm H}}{\gamma_{\rm H\beta}n_e}\left(\frac{v_{\rm out}}{R_{\rm out}}\right)L_{\rm H\beta,broad} 10^{0.4A_{\rm H\beta}},
    \label{eq:mass_outflow_rate}
\end{equation}
where $m_{\rm H}$ is the atomic mass of hydrogen.  $\gamma_{\rm H\beta}$ is the H$\beta$ emissivity for case B recombination with assumed electron temperature ${T_e=10^4}$~K \citep[$\gamma_{\rm H\beta}=1.24\times10^{-25}$~erg~cm$^3$~s$^{-1}$;][]{osterbrock2006astrophysics}.  $n_e$ is the local electron density in the outflow, where we assume $n_e=100$~cm$^{-3}$ \citepalias{reichardtchu2022spatiallyresolved}.  $v_{\rm out}$ is the maximum outflow velocity found from fitting the emission lines, here we use the $v_{\rm out}$ from H$\beta$.  $R_{\rm out}$ is the radial extent of the outflow, where we assume $R_{\rm out}=0.5$~kpc.  $L_{\rm H\beta}$ is the H$\beta$ luminosity from the fitted broad Gaussian component. Finally, $A_{\rm H\beta}$ is the extinction derived from the observed H$\beta$/H$\gamma$ ratio for the narrow line component and a \citet{calzetti2001dustopacity} attenuation curve (see Section~\ref{subsec:cont_subtract}), which is discussed further below.  

Any calculation of $\dot{M}_{\rm out}$ heavily depends on the assumption made for the geometry and electron density of the outflow. We have assumed $R_{\rm out}=0.5$~kpc and $n_e=100$~cm$^{-3}$ for direct comparison to our previous results \citetalias{reichardtchu2022resolvedmaps, reichardtchu2022spatiallyresolved}.  In addition, 0.5~kpc is similar to the resolution of our KCWI spaxels when averaged across our sample of 10 galaxies. The true value of $R_{\rm out}$ may be anywhere from $r_{90}$ for gas that has travelled far from its launching site, to only a few hundred parsecs for gas that has only just been driven out of the galaxy disk. 
For a full discussion of the motivation and likely systematic uncertainty on $R_{\rm out}$ see \citetalias{reichardtchu2022resolvedmaps}; for a discussion  of the choices for $\gamma_{\rm H\beta}$ and $n_e$ see \citetalias{reichardtchu2022spatiallyresolved}.  The impact of the assumptions for these values on is explored more fully in Section~\ref{subsec:mass_loading_factor} below.


\begin{figure*}
    \centering

    \includegraphics[width=\textwidth]{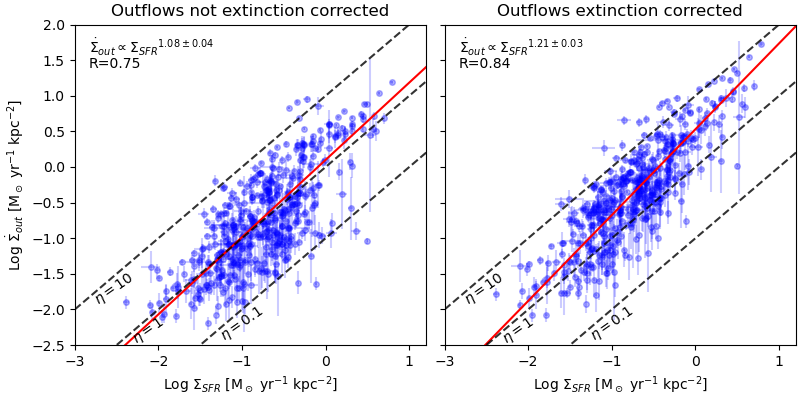}

    \caption{The ionised gas outflow mass flux, $\dot{\Sigma}_{\rm out}$, compared to the SFR surface density, $\Sigma_{\rm SFR}$, for sub-kpc resolution spaxels in local starbursting disk galaxies with evidence for outflow components.  
    We fit a relationship between the two quantities using the method of orthogonal distance regression, which returns a nonlinear correlation, given with the Pearson correlation coefficient in the top left corner and plotted as a red solid line.  
    Dashed black lines show constant mass loading factors. The left panel shows no extinction correction on the outflow. The right panel shows $\dot{\Sigma}_{\rm out}$ with the same extinction correction applied as for gas in the disk. }
    \label{fig:sigsfr_mout_allgals}
\end{figure*}

We do not have sufficient S/N on H$\gamma$ to resolve the outflow component, and then determine an extinction correction for the outflow. We note that it is often an implicit assumption that $\dot{M}_{\rm out}$ includes the same extinction as the ISM, for example in calculating $\eta \propto F_{\rm broad}/F_{\rm narrow}$. While dust certainly exists in outflows \citep[e.g.][]{engelbracht2006extendedmIR_M82, roussel2010cooldust_M82}, the actual value of extinction and how that corresponds with extinction in the midplane of the galaxy is unclear. 
\citet{xu2022CLASSYproperties} approximated the total dust extinction using absorption-line outflows of CLASSY galaxies and found that most of the dust responsible for extinction resides within the static ISM.  Conversely, \citet{avery2021incidence} found a Balmer decrement for the majority of their outflows consistent within the errors with the ISM gas, with a skew towards significantly enhanced extinction in the outflowing gas. We, therefore, compare our results assuming the same extinction correction that we applied for the disk to the outflow, to our results when assuming no extinction in the outflow. 
The actual value of extinction may lie in between these two extremes, and may vary from galaxy to galaxy.

The outflow mass flux is the mass outflow rate divided by the surface area of the measurement $\Dot{\Sigma}_{\rm out}=\Dot{M}_{\rm out}/{\rm Area}$.  The outflow mass flux is a more useful measurement in resolved outflow studies than the outflow mass rate, and can be compared to results from simulations, especially high-resolution box simulations \citep[e.g.][]{kim2017tigress,rathjen2023silccVIIgaskinematics}.

In the right-hand panel of Fig.~\ref{fig:sigsfr_mout_allgals} we plot the extinction-corrected outflow mass flux against the SFR surface density for all galaxies in our sample.  We measure ${\dot{\Sigma}_{\rm out}\sim0.01~M_\odot}$~yr$^{-1}$~kpc$^{-2}$ to $\sim100~M_\odot$~yr$^{-1}$~kpc$^{-2}$ over three orders of magnitude in $\Sigma_{\rm SFR}$.  
The median $\log(\dot{\Sigma}_{\rm out} [M_\odot$~yr$^{-1}$~kpc$^{-2}])$ for our sample was -0.34 with a 1$\sigma$ range of $\pm1.0$~dex.    
We find a Pearson correlation coefficient of $R_P=0.84$ ($p-$value$\ll10^{-3}$) 
for $\log\dot{\Sigma}_{\rm out}-\log\Sigma_{\rm SFR}$.  Due to the large range of error bars in outflow mass flux, fitting the data would be biased towards the high S/N points at high SFR surface density.  We, therefore, constrain the error bars to have a minimum uncertainty of 0.15~dex in $\dot{\Sigma}_{\rm out}$ for the fit.  Fitting the data using the method of orthogonal distance regression, we find a superlinear relationship for the extinction-corrected outflows of 
\begin{equation}
    \dot{\Sigma}_{\rm out} = 10^{0.53\pm0.03}~\Sigma_{\rm SFR}^{1.21\pm0.03}.
    \label{eq:sigmaout_sigmasfr}
\end{equation}
We show this fit as a red line in the right-hand panel of Fig.~\ref{fig:sigsfr_mout_allgals}. This is slightly steeper than 
the relationship found in \citetalias{reichardtchu2022spatiallyresolved} ($\dot{\Sigma}_{\rm out}\propto\Sigma_{\rm SFR}^{1.06\pm0.10}$).  When the errorbars are used without modification, we find a steeper correlation $\dot{\Sigma}_{\rm out}\propto\Sigma_{\rm SFR}^{1.29\pm0.03}$, which is consistent with the relationship found with unconstrained errorbars in \citetalias{reichardtchu2022spatiallyresolved}.   We observe values of $\dot{\Sigma}_{\rm out}$ an order of magnitude lower than found in \citetalias{reichardtchu2022spatiallyresolved}, and these low S/N results may weight the fit towards low $\dot{\Sigma}_{\rm out}$ values when the error bars are unconstrained.

In the left panel of Fig.~\ref{fig:sigsfr_mout_allgals}, we show the same relationship if we assume no extinction in the outflow. With no extinction correction, we find a median $\log(\dot{\Sigma}_{\rm out} [M_\odot$~yr$^{-1}$~kpc$^{-2}])$ of -0.78 with a 1$\sigma$ range of $\pm1$~dex.    
We also find a slightly shallower relationship between the mass outflow flux and the SFR surface density, such that $\Dot{\Sigma}_{\rm out}\propto\Sigma_{\rm SFR}^{1.08\pm0.04}$.  This is consistent with the relationship found in \citetalias{reichardtchu2022spatiallyresolved} within uncertainties.  
If we do not apply an extinction correction to the outflows in the same manner as we do to the SFR, we find mass loading factors a factor of $\sim$3 lower, with a median mass loading factor, $\eta$, of 1.04 with a 1$\sigma$ range of $\pm0.48$~dex.  
We use the non-extinction corrected values of $\dot{\Sigma}_{\rm out}$ for the remainder of the paper. 



In Appendix~\ref{appendix:ind_gals} we show the same relationship as Fig.~\ref{fig:sigsfr_mout_allgals}, but plotting each galaxy in our sample individually.  We find that each galaxy has a trend of increasing outflow mass flux with increasing SFR surface density. We interpret this to suggest that the important change from galaxy to galaxy is the overall degree of star formation, yet roughly the same local relationship between  $\dot{\Sigma}_{\rm out}$ and $\Sigma_{\rm SFR}$ remains.

There is a wide range of results in the literature for the observed relationship between the total outflow mass rate and SFR using galaxy-integrated and stacked resolved measurements.  \citet{fluetsch2019coldmolecularoutflows} found $\dot{M}_{\rm out}\propto$~SFR$^{1.19\pm0.06}$ for molecular gas outflows in local star-forming galaxies. 
\citet{avery2021incidence} found a shallower relationship, $\dot{M}_{\rm out}\propto{\Sigma_{\rm SFR}}^{0.25\pm0.09}$, or $\dot{M}_{\rm out}\propto{\rm SFR}^{0.84\pm0.08}$, for ionised gas outflows using radially binned star-forming galaxies from MaNGA, excluding AGN from their sample.  In a follow-up paper of the same galaxies, \citet{avery2022cooloutflowsMaNGA} found a relationship of $\dot{M}_{\rm out}\propto{\rm SFR}^{0.74\pm0.2}$ for the neutral gas outflows.  
For a sample of dwarf starbursting galaxies from the DWALIN survey, \citet{marasco2023shakennotexpelled} found an even shallower result, $\dot{M}_{\rm out}\propto$~SFR$^{0.4}$.  However, we caution that these galaxy-averaged measurements almost certainly include regions that do not drive an outflow but do contribute to the total SFR.


Using magnetohydrodynamic box simulations of the ISM and star formation-driven outflows with the SILCC framework, \citet{rathjen2023silccVIIgaskinematics} found $\dot{\Sigma}_{\rm out}\propto\Sigma_{\rm SFR}^{0.81\pm0.12}$ for the total gas outflow mass flux.  They suggested that more massive galaxies with higher $\Sigma_{\rm SFR}$ have more difficulty driving outflows from their larger potential wells, leading to a sublinear correlation.  
Considering only the warm ionised gas, \citet{rathjen2023silccVIIgaskinematics} found a steeper relationship of $\dot{\Sigma}_{\rm out}\propto\Sigma_{\rm SFR}^{1.40\pm0.24}$.  This is steeper than the slope which we fit in Fig.~\ref{fig:sigsfr_mout_allgals} (see also Eq.~\ref{eq:sigmaout_sigmasfr}).  However, our galaxies cover a range of SFR surface densities that extends at least an order of magnitude beyond the maximum SFR surface density ($\Sigma_{\rm SFR}\sim0.2~M_\odot$~yr$^{-1}$~kpc$^{-2}$) modelled by \citet{rathjen2023silccVIIgaskinematics}, and so a complete comparison is difficult.
It is possible that the less dense ionised gas can be accelerated more efficiently than the cold phases of gas, leading to a steeper correlation.  
This difference could explain the shallower slope which \citet{avery2022cooloutflowsMaNGA} observed for their molecular gas outflows than we find for ionised gas outflows, although \citet{avery2022cooloutflowsMaNGA} found agreement at the $1\sigma$ level between their observations of ionised and neutral gas outflows.

\subsection{Mass loading factor}
\label{subsec:mass_loading_factor}

The mass loading factor describes the efficiency of a galaxy in driving outflows.  The resolved mass loading factor is the ratio between the mass outflow flux and the star formation rate surface density
\begin{equation}
    \eta = \frac{\dot{\Sigma}_{\rm out}}{\Sigma_{\rm SFR}}.
    \label{eq:mlf}
\end{equation}
The mass loading factor relates the rate of gas leaving a region of the galaxy to the underlying star formation driving the outflow.  Regions of a galaxy with a mass loading factor greater than 1 are driving more gas out of the galaxy than they are turning into stars.  

In 
Fig.~\ref{fig:sigsfr_mout_allgals} we show lines of constant mass loading factor as diagonal dashed black lines.  When we consider all galaxies in our sample, we find that for the majority of spaxels we measure an ionised gas mass loading factor between 0.1 and 10.  We find a median mass loading factor of 1.04 with a 1$\sigma$ scatter of $\pm0.48$~dex.  We did not find a statistically significant correlation between the mass loading factor and the SFR surface density, with a Pearson correlation value of $R_P=-0.04$ ($p$-value$=0.35$) between $\log(\eta)$ and $\log(\Sigma_{\rm SFR})$.   
This is expected due to the almost linear correlation we find between the mass outflow flux and the SFR surface density.
Our results suggest that in the majority of the regions where we observe an outflow, the star formation is efficiently coupled to the gas to drive the outflow.

Outflows are by nature multiphase. When calculating the mass loading factor we, therefore, need to take into account the total mass budget of the outflow.
\citet{fluetsch2019coldmolecularoutflows} found roughly comparable mass outflow rates from ionised and molecular gas observations of outflows in four galaxies with a similar specific star formation rates, sSFR$=$SFR/$M_*$, to the galaxies in our sample.  This would imply a total mass loading factor for our galaxies of order $\eta\sim2-20$. In contrast to this, 
\citet{fluetsch2021propertiesmultiphase} found that for local ULIRGs, the mass of ionised gas in the galaxy wind is a very small fraction ($<5\%$) of the total mass outflow, with neutral gas making up $\sim10\%$ and molecular gas up to $\sim95\%$ of the outflowing mass.  Similarly, \citet{robertsborsani2020outflowsMANGA} found that ions contribute $<1\%$ of the mass to the total outflowing gas mass.    
Following these studies, the total mass loading factor for the galaxies we observe here could be anywhere from $\eta\sim10-100$ or even greater.

\citet{avery2022cooloutflowsMaNGA} compared the analysis of ionised gas outflows in the MaNGA galaxies from \citet{avery2021incidence} to observations of neutral gas outflows studied using the Na~I~D absorption feature.
\citet{avery2022cooloutflowsMaNGA} accounted for the dust extinction in the outflow separately from that within the ISM of the galaxy and found that this increased the fraction of the total gas mass measured in ionised gas.  Even with this correction, the ionised gas phase mass they measured remained $\sim1.2$~dex lower than the neutral gas phase mass.  They found an average neutral gas mass loading factor of order unity for non-AGN MaNGA galaxies within a similar mass range to our galaxies.  If our galaxies follow a similar trend, then we could expect to observe neutral gas mass loading factors of $\sim10-100$, which is far higher than the neutral gas mass loading factors that \citet{avery2022cooloutflowsMaNGA} found.  Alternatively, if we expect the galaxies in our sample to have similar neutral gas mass loading factors to those found by \citet{avery2022cooloutflowsMaNGA}, then it is possible that neutral gas observations of the galaxies in our sample would find roughly comparable ionised and neutral gas mass loading factors.  This would then be similar to the results from \citet{fluetsch2019coldmolecularoutflows}.

However, we urge caution when comparing observational results between resolved studies such as this one and studies utilising stacked spectra or total galaxy measurements such as \citet{avery2022cooloutflowsMaNGA} and \citet{fluetsch2019coldmolecularoutflows}.  Normalising the mass outflow rate by the star formation rate may include star formation from regions of the galaxy that do not drive an outflow in non-resolved studies.  This can result in an artificially low mass loading factor for some observations.  

Moreover, observations of the mass loading factor inherently include a large number of assumptions about the geometry and electron density of the outflow.  These assumptions can vary dramatically between studies and can have a large impact on the results.  
For example, following from Eq.~\ref{eq:mass_outflow_rate} and \ref{eq:mlf}, increasing the assumed electron density from $n_e=100$~cm$^{-1}$ to $n_e=300$~cm$^{-1}$ decreases the mass loading factor by $\sim0.5$~dex.  The electron density within the disk has been observed to depend on the local SFR surface density \citep[e.g.][]{kaasinen2017cosmosOIIsurveyelectrondensity, davies2021KMOS3Delectrondensities}. If the electron density of the outflow also depends on the SFR surface density such that the electron density decreases for brighter outflows, this would steepen the correlation between the mass loading factor and SFR surface density, bringing our result closer to predictions from simulations \citep[e.g.][]{kim2020firstresultssmaug, pandya2021characterisingFIRE2}.
Alternatively, again following from Eq.~\ref{eq:mass_outflow_rate} and \ref{eq:mlf}, increasing the assumed outflow radius from $R_{\rm out}=0.5$~kpc to $R_{\rm out}=20$~kpc 
decreases the mass loading factor by $\sim1.5$~dex.  If the $R_{\rm out}$ increases for outflows driven by stronger star formation, this might offset a change in the electron density.  It is therefore difficult to definitively say how our assumptions impact our results.  High-resolution studies of outflows from edge-on galaxies such as the MUSE/VLT GECKOS Large Program are likely to make progress on understanding these systematics (Elliot et al. {\em in prep}).

In addition, while we expect that there should be an order of magnitude more molecular gas than ionised gas in the total outflow mass budget, regions of the galaxy with higher SFR surface densities may well be ionising more gas, increasing the fraction of ionised gas in the outflow.  Simply assuming the same ratio of ionised-to-molecular gas in the outflow for all SFR surface densities may therefore lead us to overestimate the total mass loading factor.  For example, in the MHD stratified galaxy patch simulation SILCC, \citet{rathjen2023silccVIIgaskinematics} found a steeper relationship between the warm and ionised outflow mass rate and the SFR surface density than between the total outflow mass rate and the SFR surface density.  This suggests that for regions with higher SFR surface densities, ionised gas may make up a larger fraction of the total outflowing mass.



Simulations use the relationship between the mass loading factor and the SFR surface density to test the driving mechanisms of outflows.  
A number of simulations predict a negative relationship of $\eta_{\rm ion}$ with SFR surface density \citep{fielding2017SNelaunch, li2017supernovaedriven, kim2020firstresultssmaug, pandya2021characterisingFIRE2}. With our sample of galaxies, however, we find a very shallow positive relationship between the mass loading factor and the SFR surface density. 
We now compare the mass loading factor values from the simulations to those measured in our observations.
Using the FIRE suite of cosmological zoom-in simulations, \citet{anglesalcazar2017cosmicbaryoncycle_FIRE} found total gas mass loading factors of $\eta\sim2$ for galaxies in a similar galaxy mass range to our sample.
From the updated FIRE-2 simulations, \citet{pandya2021characterisingFIRE2} found that the warm ($10^3~{\rm K}<T<10^5$~K) gas represents less than 10\% of the total mass loading for galaxies with a similar stellar mass to our sample. For $\Sigma_{\rm SFR}\sim0.1-10~M_\odot$~yr$^{-1}$~kpc$^{-2}$ they find warm gas mass loading factors ranging from $\eta\sim0.04-10$.  These values are within the range of the mass loading factors we find for the ionised gas phase observations in our galaxies. 
Using the SMAUG-TIGRESS simulations, \citet{kim2020firstresultssmaug} similarly found mass loading factors for the cool ($T\sim10^4$~K) gas of $\eta\sim0.2-10$ for $\Sigma_{\rm SFR}\sim0.1-1~M_\odot$~yr$^{-1}$~kpc$^{-2}$.  These values are consistent with the values we observe in our galaxies at a similar SFR surface density.

We note that we are not comparing apples with apples.  \citet{kim2020firstresultssmaug} simulate solar neighbourhood-like conditions and do not reach the high $\Sigma_{\rm SFR}$ environments of some regions of our targets.  In addition, \citet{pandya2021characterisingFIRE2} calculate total values of $\eta$ and SFR surface density for entire galaxy haloes, while we calculate these values for resolved regions.  While our extended sample covers a larger range of SFR surface densities than \citetalias{reichardtchu2022spatiallyresolved}, our results are however still within the scatter of results returned for the warm ionised gas by \citet{pandya2021characterisingFIRE2}.

\subsection{Stellar mass surface density}

\begin{figure*}
    \centering
    \includegraphics[width=\textwidth]{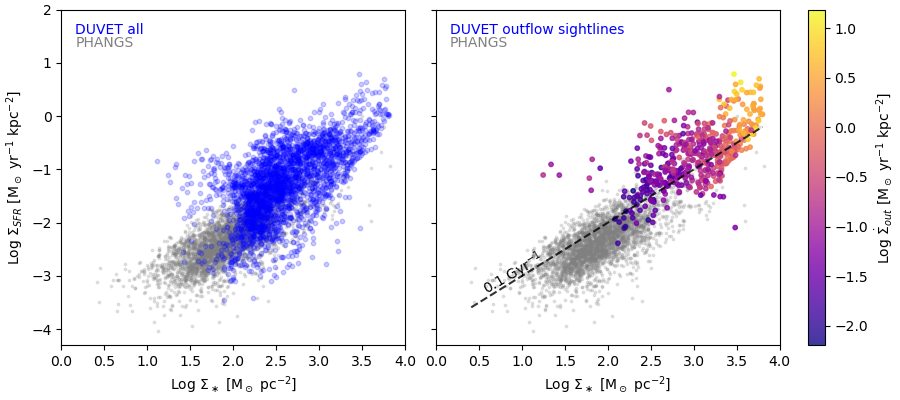}
    \caption{The so-called ``resolved main-sequence'' for DUVET galaxies. The left panel shows all spaxels within $r_{90}$. The right panel shows the same region, but only for those spaxels with outflows. Grey points are kiloparsec-scale measurements from the PHANGS sample. The line is an {\em ad hoc} seperation showing that outflows become more common for $\Sigma_{\rm SFR}/\Sigma_\ast > 0.1$~Gyr$^{-1}$. Outflow points are coloured by outflow mass flux. }  
    \label{fig:resolved_ms}
\end{figure*}

\begin{figure*}
    \centering
    \includegraphics[width=\textwidth]{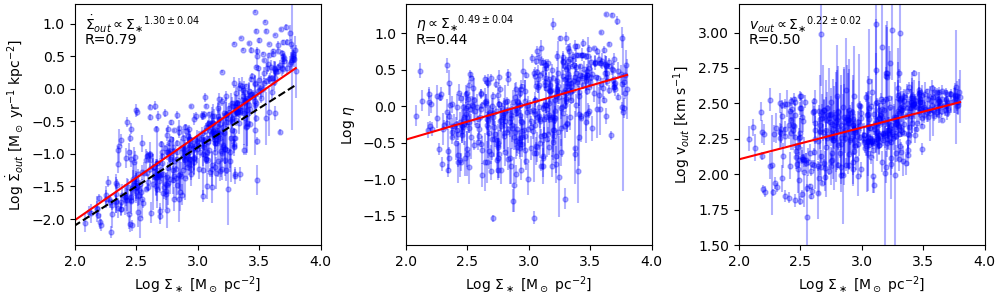}
    \caption{The relationship between the resolved stellar mass surface density and the outflow mass flux (left), the mass loading factor (middle), and the outflow velocity (right).  We fit a relationship between the two quantities for each panel using the method of orthogonal distance regression.  The resulting nonlinear correlations are given with the Pearson correlation coefficient in the top left corner of each panel, and are plotted as solid red lines. In the left panel, the dashed black line shows $\dot{\Sigma}_{\rm out}\propto\Sigma_\ast^{1.2}$, which would be the case if the $\dot{\Sigma}_{\rm out}-\Sigma_\ast$ relation followed the $\dot{\Sigma}_{\rm out}-\Sigma_{\rm SFR}$ relation.}
    \label{fig:stellarmass_outflow}
\end{figure*}

In Fig.~\ref{fig:resolved_ms} we show the so-called ``resolved main-sequence'' relationship between $\Sigma_{\rm SFR}$ and $\Sigma_\ast$ for DUVET targets in this work. We derive stellar masses using 2MASS H-band observations, which are available for all of our sources. The 2MASS data is convolved and resampled to match the DUVET KCWI images. Our galaxies are selected to be starbursts, so to convert to stellar mass we assume an LMC mass-to-light ($M/L$) ratio \citep{eskew2012spitzerfluxtomass} of $M/L_H = 0.11$ by scaling the 3.6~$\mu$m M/L to H-band, and the \citet{salpeter1955imf} IMF to a \citet{kroupa2003imf} IMF. 
We note that we tested the \citet{meidt2012reconstructingstellarmass_PAH} method using Spitzer 3.6~$\mu$m and 4.5~$\mu$m images, and determined that Spitzer 3.6~$\mu$m images in our sample are likely to be very strongly affected by 3.3~$\mu$m PAH feature, which varied significantly across our targets. We, therefore, opted not to use methods like \citet{meidt2012reconstructingstellarmass_PAH} as the PAH feature introduced extra systematics to the conversion to stellar mass. In Fig.~\ref{fig:resolved_ms} we compare our $\Sigma_{\rm SFR}$ and $\Sigma_\ast$ to those of the PHANGS sample \citep{sun2023starformationlaws}, measured with comparable resolution, and find that for regions of DUVET galaxies with $\Sigma_{\rm SFR}$ that is similar to values of PHANGS targets the $\Sigma_\ast$ is likewise comparable. 

The DUVET sample is selected to have global SFR and M$_*$ that is nominally at least 5$\times$ higher than the star-forming main sequence, and this is reflected in Fig.~\ref{fig:resolved_ms}. In the left panel of the figure, we show all lines of sight, the majority of which have higher $\Sigma_{\rm SFR}/\Sigma_\ast$ than PHANGS galaxies, which are more representative of the resolved star-forming main sequence. Overall, the DUVET targets have a spread in  $\Sigma_{\rm SFR}/\Sigma_\ast$  that begins at the resolved main sequence and skews upwards. 

In the right panel of Fig.~\ref{fig:resolved_ms} we show only those lines of sight in which our automatic detection method identifies outflows. Nearly 100\% of the lines of sight containing outflows are above the main sequence. We also plot a line corresponding to $\Sigma_{\rm SFR}/\Sigma_\ast = 0.1$~Gyr$^{-1}$. Outflows are very rare below this specific star formation rate, at any value of $\Sigma_\ast$. We find that below $\Sigma_{\rm SFR}/\Sigma_\ast = 0.1$~Gyr$^{-1}$ only 12\% of all spaxels within $r_{90}$ are determined to have outflows, and above this value 20\% do have outflows. Out of all the spaxels with outflows, 72\% have $\Sigma_{\rm SFR}/\Sigma_\ast > 0.1$~Gyr$^{-1}$.

A large number of studies conclude that outflows are more common at higher specific SFR \citep[e.g.][]{chen2010absorption, Rubin2014evidence, forsterschreiber2019kmos3d,avery2021incidence,robertsborsani2020outflowsMANGA,veilleux2020cooloutflows}. Our result in  Fig.~\ref{fig:resolved_ms} shows that this extends down to sub-kpc scales. More specifically, \cite{forsterschreiber2019kmos3d} showed that $z\approx2$ KMOS3D galaxies that are above the main sequence are much more likely to host outflows, similar to our result. There are both theoretical and observational arguments \citep{murray2011radiationpressure,heckman2002extragalactic} that a minimum $\Sigma_{\rm SFR}\gtrsim 0.1$~M$_{\odot}$~yr$^{-1}$~kpc$^{-2}$ is needed to drive observable outflows.
While we do find outflows become less common below this threshold, it is clear from Figures~\ref{fig:sigsfr_mout_allgals} and \ref{fig:resolved_ms} that they are still present and seem to extend the correlations of higher $\Sigma_{\rm SFR}$ winds.  We find that 33\% of the spaxels we observe to contain outflows fall below this threshold.  
It is possible that any threshold value in $\Sigma_{\rm SFR}$ is likely to increase at higher $\Sigma_\ast$.  This can be understood by a simple argument in which the higher $\Sigma_\ast$ exerts a greater gravitational force on the outflow gas. The gas may require more energy, from higher $\Sigma_{\rm SFR}$, to escape the local potential. This would be consistent with where we find outflows in the DUVET targets.  

In Fig.~\ref{fig:stellarmass_outflow} we compare $\Sigma_\ast$ for individual spaxels to the outflow properties ($\dot{\Sigma}_{\rm out}$, $\eta$ and $v_{\rm out}$). All three show positive correlations. The correlation of $\dot{\Sigma}_{\rm out}-\Sigma_\ast$ is similarly strong as that of $\dot{\Sigma}_{\rm out}-\Sigma_{\rm SFR}$, with a Pearson correlation coefficient $R_P=0.84$ ($p-$value$\ll10^{-3}$).  
We find, however, that the powerlaw of this relationship is steeper, such that $\dot{\Sigma}_{\rm out}\propto\Sigma_\ast^{1.45\pm0.04}$, where the powerlaw index scaling with $\Sigma_{\rm SFR}$ is closer to $\sim1.2$. 
Similarly, the correlation of $\eta-\Sigma_\ast$ is steeper than $\eta-\Sigma_{\rm SFR}$, such that $\eta \propto \Sigma_\ast^{0.55\pm0.05}$, with a Pearson correlation coefficient of $R_P=0.45$ ($p-$value$\ll10^{-3}$). 
The powerlaw index scaling for $\eta-\Sigma_{\rm SFR}$ is closer to zero.
The correlation of $v_{\rm out}$ with $\Sigma_\ast$ is shallow and similar to that with $\Sigma_{\rm SFR}$. 

We do not know why the correlation of $\dot{\Sigma}_{\rm out}$ with $\Sigma_\ast$ should be so steep. For spaxels with outflows the correlation of $\Sigma_{\rm SFR}$ with $\Sigma_\ast$ is roughly consistent with linear. Therefore, if we assume that   $\dot{\Sigma}_{\rm out} - \Sigma_\ast$ is a secondary correlation between $\dot{\Sigma}_{\rm out}$ and $\Sigma_{\rm SFR}$ then we would expect the same powerlaw. 

We note that the steep powerlaw of $\dot{\Sigma}_{\rm out} \propto \Sigma_*^{1.45}$ may be a relic of systematic uncertainties. We assumed a constant $M/L$ across our sample. However, if this changes due to the local stellar populations or position in the galaxy, then this could impact the powerlaw. In this case, however, the most likely scenario would be that higher $\Sigma_{\rm SFR}$ would have younger stellar populations and thus lower $M/L$. This would steepen the relationship.

There are also physical arguments for a steeper relationship with $\Sigma_\ast$. For example, higher $\Sigma_\ast$ implies a stronger local gravitational potential. In principle, this requires outflows to be stronger for them to overcome the local gravity, and be observed in our sample. Secondly, as the highest $\Sigma_\ast$ spaxels are preferentially in the galaxy center, it may be that gas is preferentially built up in these regions and thus able to drive stronger winds. 

\subsection{Outflow covering fraction}

The covering fraction of outflows is defined by the total fraction of starlight that is covered by outflowing gas.  This quantity is necessary when calculating the mass-outflow rate for entire galaxy outflow measurements \citep[e.g. at higher redshift;][]{Rubin2014evidence, davies2019kiloparsec} or for edge-on galaxy outflow measurements. In absorption line measurements it is typically inferred from the depth of the absorption line. Our resolved outflow measurements allow us to measure this directly via geometry. See \cite{mcpherson2023duvet_mrk1486} for a similar calculation in an edge-on outflow galaxy. 

\begin{figure*}
    \centering
    \includegraphics[width=0.9\textwidth]{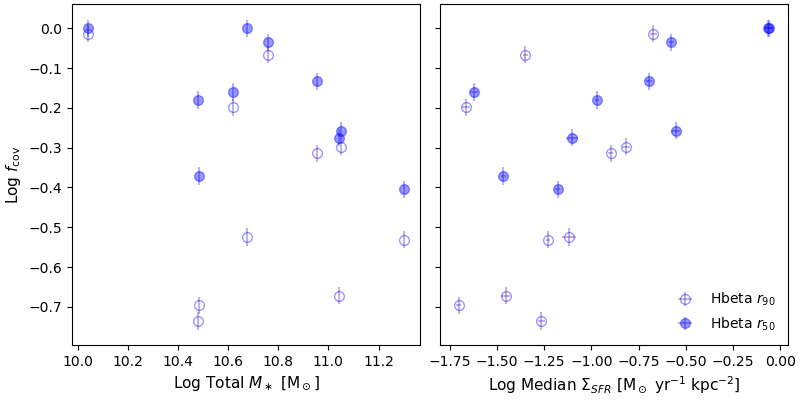}
    \caption{The fraction of spaxels with outflows detected in in H$\beta$ within $r_{50}$ (coloured points) and $r_{90}$ (open points) for each galaxy is plotted against the total stellar mass (left panel), and the median SFR surface density (right panel) of the galaxies.   
    We find the strongest correlation is a dependence of the covering fraction on the median SFR surface density.}
    \label{fig:covering_frac}
\end{figure*}

Here we calculate the covering fraction as  $f_{\rm cov} \equiv N_{\rm outflow}/N_{\rm total}$, where $N_{\rm outflow}$ and $N_{\rm total}$ are, respectively, the number of spaxels with an outflow and the total number of spaxels. 
We calculate this for each galaxy within the 90\% starlight radius  ($r_{90}$) and half-light radius ($r_{50}$).  In Fig.~\ref{fig:covering_frac} we compare $f_{\rm cov}$ within $r_{50}$ and $r_{90}$ to the total galaxy stellar mass, 
and the median SFR surface density within $r_{50}$ and $r_{90}$ for outflows detected in H$\beta$.

Across our galaxy sample, we find a possible trend of decreasing covering fraction with increasing stellar mass, shown in the left panel of Fig.~\ref{fig:covering_frac}.  The correlation between the covering fraction and the total galaxy stellar mass has a Pearson correlation coefficient of $R_P=-0.37$ ($p-$value$=0.11$) for the relationship $\log(f_{\rm cov})-\log(M_*)$. For galaxies with $\log M_*[M_\odot]=10.7-11.5$ we find a median covering fraction of 49\% within $r_{90}$.  For the lower mass galaxies in the sample ($\log M_*[M_\odot]=10-10.7$), we find the covering fraction increases to 30\%.  The decrease in the fraction of outflow spaxels with increasing stellar mass seems connected to our result in Fig.~\ref{fig:resolved_ms}, where individual spaxels with higher $\Sigma_\ast$ require higher $\Sigma_{\rm SFR}$ to be observed with an outflow. Similarly, this can be understood given that galaxies with a higher stellar mass have a larger gravitational potential well, making it harder to accelerate outflows to the velocities necessary to leave the disk. 

The right panel of Fig.~\ref{fig:covering_frac} shows the relationship between the covering fraction and the median value of the SFR surface density. This is calculated within $r_{50}$ to compare to the covering fraction within $r_{50}$, and similarly within $r_{90}$ to compare to the covering fraction within $r_{90}$.  We find an overall trend of increasing covering fraction with increasing median SFR surface density. This trend has a higher statistical significance, with a Pearson correlation coefficient of $R_P=0.62$ ($p-$value$=0.004$) for $\log(f_{\rm cov})-\log(\Sigma_{\rm SFR, median})$. 
Given the local correlations between outflow properties and $\Sigma_{\rm SFR}$, this trend seems straightforward to understand. If the typical $\Sigma_{\rm SFR}$ in a galaxy is larger then this galaxy will have a larger area in which outflows are present. 

We also test the correlation of the covering fraction with stellar mass and median SFR surface density for our results using [OIII]~$\lambda5007$.  We find a weaker negative correlation with stellar mass (Pearson correlation coefficent $R_P=-0.28$, $p-$value$=0.23$), and a similar correlation with median SFR surface density (Pearson correlation coefficent $R_P=0.65$, $p-$value$=0.001$).  However, overall fewer spaxels contain outflows.


Additionally, we also test the change in covering fraction from $r_{50}$ out to $r_{90}$.  
We find that the fraction of spaxels with outflows decreases between $r_{50}$ and $r_{90}$ by a median of 0.24~dex for H$\beta$ and 0.25~dex for [OIII]~$\lambda5007$, suggesting that the majority of outflows are centrally located, although not all outflows are launched from the centre of the galaxies.  This is similar to results from \citet{avery2022cooloutflowsMaNGA} for the MaNGA galaxies, who found that most neutral gas outflows are concentrated centrally within $\sim0.25~R_e$. 

Using absorption line measurements for 105 galaxies at $0.3<z<1.4$, \citet{Rubin2014evidence} found a median covering fraction of $\sim0.65$ for both the Mg~II and Fe~II absorption lines. This is far higher than what we find using emission lines to trace the ionised gas.  However, absorption line measurements are sensitive to gas clouds at any distance along the line-of-sight to the host galaxy.  It is likely that they are able to probe outflowing gas that has reached further from the galaxy than our emission line measurements, which probe the base of the outflow.  Their higher covering fractions could, therefore, be caused by outflowing gas which has either been driven by underlying star formation which is no longer observable, or has expanded further across the face of the galaxy due to the outflow opening angle.  

In addition, \citet{Rubin2014evidence} found covering fractions $\gtrsim0.5$ in 75\% of the absorption lines where they also measured equivalent widths for the outflow component EW$^{16\%}_{\rm flow}>0.2$~\AA.  They found a dependence of the outflow equivalent width on the covering fraction.  They also found that the outflow equivalent width was correlated with the SFR, such that galaxies with a higher SFR launch more material with a larger range of velocities.  
\citet{prusinski2021connectinggalacticoutflows} also found a relationship between the equivalent width and the covering fraction such that galaxies with more intense star formation drive outflows with a higher covering fraction and a wider range of velocities.
These studies are in agreement with our result that galaxies with a higher SFR surface density have a higher covering fraction.



\subsection{Comparison to total galaxy measurements}
\label{subsec:total_gal}

\begin{figure*}
    \centering
    \includegraphics[width=0.8\textwidth]{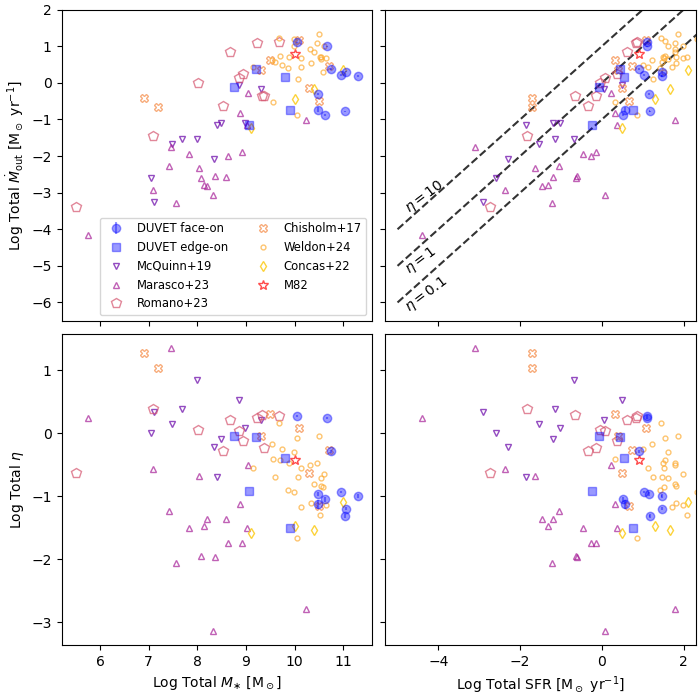}
    \caption{Summed total galaxy quantities for our 10 face-on DUVET galaxies (blue circles) and 5 DUVET edge-on galaxies (blue squares, McPherson et. al. \textit{in prep}) compared with entire galaxy measurements from nearby dwarf galaxies measured in emission \citep[down triangles, up triangles:][respectively]{mcquinn2019galacticwinds, marasco2023shakennotexpelled}, atomic gas outflows in nearby dwarf galaxies measured in [CII]~158~$\mu$m emission \citep[pentagons:][]{romano2023sfdrivenoutflows}, nearby star-forming galaxies measured in absorption \citep[crosses:][]{chisholm2017mass}, main-sequence star-forming galaxies at cosmic noon measured in emission \citep[diamonds, open circles:][respectively]{concas2022KLEVERionisedoutflows, weldon2024MOSDEFionizedoutflows}, and nearby starburst M~82 \citep[red star:][]{greco2012massmeasurement_M82, xu2023radialdistributions_M82}.  The panels show total galaxy measurements for mass outflow rate $\dot{M}_{\rm out}$ against total stellar mass $M_\ast$ (top left), and total star formation rate (top right); and the mass loading factor $\eta$ against total stellar mass $M_\ast$ (bottom left), and total star formation rate (bottom right).  Dashed lines in the top right panel show lines of constant mass loading factor.  For clarity, only the 10 face-on DUVET targets studied in this paper are shown with error bars. Our DUVET targets fall within the expected scatter for total galaxy measurements.  The outlier from the DUVET face-on galaxy sample is UGC~12150, which we found had a very weak outflowing ionised gas signal.}
    \label{fig:total_gals}
\end{figure*}

In Fig.~\ref{fig:total_gals} we compare our summed total galaxy results to literature values for total galaxy measurements of mass outflow rate, mass loading factor, total stellar mass and star formation rate.  Here we use the total IR star formation rates computed in Section~\ref{subsec:duvet} using WISE photometry for our 10 galaxies (see Column~5 in Table~\ref{tab:gal_params}). We compare our local starbursting disk galaxies to three samples of nearby dwarf galaxies \citep{mcquinn2019galacticwinds, marasco2023shakennotexpelled, romano2023sfdrivenoutflows}, a sample of nearby star-forming galaxies extending from dwarfs to Milky-Way mass galaxies \citep{chisholm2017mass}, two samples of star-forming galaxies at cosmic noon \citep{concas2022KLEVERionisedoutflows, weldon2024MOSDEFionizedoutflows}, and the well-studied local starbursting system M~82 \citep[e.g.][]{greco2012massmeasurement_M82, xu2023radialdistributions_M82}.  We have also included results from 5 DUVET edge-on outflow systems (McPherson et al. \textit{in prep.}).
We note that a variety of different methods were used within the literature to obtain these values, causing scatter.  For example, \citet{mcquinn2019galacticwinds}, \citet{marasco2023shakennotexpelled}, \citet{concas2022KLEVERionisedoutflows} and \citet{weldon2024MOSDEFionizedoutflows} all used emission lines to trace the ionised gas outflows where \citet{chisholm2017mass} used absorption line measurements; and  \citet{romano2023sfdrivenoutflows} used emission lines to trace atomic gas outflows.  Nevertheless, our DUVET targets fall within the range covered by the literature values for entire galaxies.   
Total galaxy outflow measurements for our 10 face-on DUVET galaxies can be found in Table~\ref{tab:individual_gals}.

\section{Conclusions}
\label{sec:conclusions}

In this paper we present sub-kpc spatially-resolved scaling relations for star formation-driven outflows and properties of their host galaxies on 10 galaxies from the DUVET sample. We fit multi-component Gaussians to the H$\beta$ and [OIII]~$\lambda5007$ emission lines in 10 near-to-face-on galaxies.  
We find the following main results.
\begin{enumerate}
    \item We find shallow relationships of the maximum outflow velocity with the SFR surface density for both H$\beta$ ($v_{\rm out, H\beta}\propto\Sigma_{\rm SFR}^{0.20\pm0.01}$) and [OIII]~$\lambda5007$ ($v_{\rm out, OIII}\propto\Sigma_{\rm SFR}^{0.12\pm0.01}$).  These fitted relationships are not consistent with models of outflows that are dominated by momentum-driven winds \citep[$v_{\rm out}\propto\Sigma_{\rm SFR}^2$;][]{murray2011radiationpressure}, and suggests supernovae as the dominant energy source driving the wind \citep[e.g.][]{kim2020firstresultssmaug}. 

    \item We fit a nonlinear relationship between the mass outflow flux and the SFR surface density such that $\dot{\Sigma}_{\rm out}\propto\Sigma_{\rm SFR}^{1.21\pm0.03}$. 

    \item We find ionised gas mass loading factors between $\sim0.1$ and $\sim10$, and approximate a total median mass loading factor of $\sim2-100$.  We fit an almost flat relationship between the mass loading factor and the SFR surface density which has no significant correlation.

    \item We make a direct comparison of outflow properties to the stellar mass surface density and specific star formation rate. Outflows are much more common for $\Sigma_{\rm SFR}/\Sigma_\ast\gtrsim 0.1$~Gyr$^{-1}$. We note that outflows are observed to lower $\Sigma_{\rm SFR}$ than previous threshold values. We also find strong, positive correlations between outflow properties and co-located $\Sigma_\ast$. In particular we find that outflow mass flux correlates with the stellar mass surface density such that $\dot{\Sigma}_{\rm out}\propto \Sigma_\ast^{1.45\pm0.04}$. 

    \item We compare the fraction of spaxels where we find evidence for outflows within $r_{50}$ and $r_{90}$ to the total galaxy stellar mass, and the median galaxy SFR surface density within $r_{50}$ and within $r_{90}$.  We find a negative correlation with stellar mass and a positive relationship with median SFR surface density.  This is consistent with the picture where galaxies of higher mass have a larger gravitational potential well, and require more concentrated star formation to drive gas out of the disk.
\end{enumerate}

The observed scaling relations of outflow properties with the SFR surface density suggest that outflows are primarily driven by energy from supernovae.  If supernovae are primarily driving the outflows, then the winds linearly correlate to the SFR surface density.  We should therefore expect that if you change the spatial distribution of the starburst then the outflow distribution will follow.  Our covering fraction results are consistent with this picture.  

This has multiple implications.  Firstly, galaxies at higher redshift may have a different outflow structure than local starburst galaxies (e.g. M~82 and NGC~253).  While local starbursts are typically concentrated in the central kiloparsec of the galaxy, galaxies at higher redshift have clumpier morphologies, which may affect the structure of the outflowing gas.  Secondly, simulations suggest that outflows that have greater covering fractions may have different energetics. \citet{schneider2018productionofcoolgas} found that altering the geometry of the simulated galaxy wind enables more gas cooling within the outflow.

Our observation that outflows are more common in higher specific SFR regions of galaxies seems straightforward to understand. Higher $\Sigma_{\rm SFR}$ provides more energy to the wind, and higher $\Sigma_\ast$ creates more gravity reducing the outflow's ability to break out of the disk. Therefore as we increase the $\Sigma_\ast$ the $\Sigma_{SFR}$ required to drive an outflow must also increase. We note that this boundary, $\Sigma_{\rm SFR}/\Sigma_\ast\sim 0.1$~Gyr$^{-1}$ is only a factor of $\sim$2$\times$ above the resolved main-sequence for similar regions \citep{sanchez2021EDGE_CALIFA}. It may be that a combination of thresholds in both SFR surface density and specific SFR is needed to predict the location of outflows. 

We do not have a direct theoretical prediction for the steep scaling relationship of $\dot{\Sigma}_{\rm out} \propto \Sigma_{\ast}^{1.45\pm0.04}$. We know that  $\Sigma_{\rm SFR}$ correlates with both outflow properties and stellar mass surface density. This may generate a secondary correlation between the outflow and the stellar mass surface density, and the steep scaling could simply be a result of uncertainties. Alternatively, it may be that the higher stellar mass demands stronger outflows for gas to break out of the plane of the disk and be observed as an outflow. More work relating the resolved stellar mass properties, both stellar populations and kinematics, with the feedback and gas properties is clearly warranted to understand this correlation. 

There are a number of limitations to our analysis. We must make many assumptions about the outflow geometry and the electron density to calculate some of these properties, particularly the mass outflow flux and the mass loading factor.  
Future work obtaining resolved observations of outflows capable of testing these assumptions is underway (Elliot et al. {\em in prep}) with the GECKOS MUSE LP \citep{vandeSande2023GECKOSsurvey_arXiv}.  It is also unclear whether the constant mass loading we observe in our sample here persists in lower SFR surface density environments.  The mass that the star formation is able to lift out of the galaxy may begin to decline more rapidly as the star formation activity decreases.

More work adding observational constraints to the total mass loading considering the multi-phase nature of outflows and its dependence on SFR surface density and geometry is needed.
In the future, with new facilities such as the ELT, we will be able to test the contribution of underlying stellar population parameters on star formation-driven feedback in starbursting galaxies on the scale of stellar clusters.

\section*{Acknowledgements}

We would like to thank the referee for their helpful comments.
Parts of this research were supported by the Australian Research Council Centre of Excellence for All Sky Astrophysics in 3 Dimensions (ASTRO 3D), through project number CE170100013. R.R.V. and K.S. acknowledge funding support from National Science Foundation Award No. 1816462. 
B.R.C. and A.F.M. acknowledge support from the Science and Technologies Facilities Council (STFC) through grant ST/X001075/1.\\
The data presented herein were obtained at the W.~M.~Keck Observatory, which is operated as a scientific partnership among the California Institute of Technology, the University of California and the National Aeronautics and Space Administration. The Observatory was made possible by the generous financial support of the W. M. Keck Foundation. Observations were supported by Swinburne Keck program 2018A\_W185. The authors wish to recognise and acknowledge the very significant cultural role and reverence that the summit of Maunakea has always had within the indigenous Hawaiian community. We are most fortunate to have the opportunity to conduct observations from this mountain.\\
This publication makes use of data products from the Two Micron All Sky Survey, which is a joint project of the University of Massachusetts and the Infrared Processing and Analysis Center/California Institute of Technology, funded by the National Aeronautics and Space Administration and the National Science Foundation.\\
This work is based in part on observations made with the Spitzer Space Telescope, which was operated by the Jet Propulsion Laboratory, California Institute of Technology under a contract with NASA.\\
This research has made use of the NASA/IPAC Extragalactic Database (NED), which is funded by the National Aeronautics and Space Administration and operated by the California Institute of Technology.\\
This work made use of the following \textsc{Python} modules: \textsc{Astropy} \citep{astropy2013, astropy2018, astropy2022}, \textsc{numpy} \citep{harris2020numpy}, \textsc{matplotlib} \citep{hunter2007matplotlib}, \texttt{cmasher} \citep{vandervelden2020cmasher}, and \texttt{threadcount} which relies heavily on \texttt{lmfit} \citep{newville2019lmfit0.9.14} and \texttt{mpdaf} \citep{mpdaf2016sourcecodelib, mpdaf2017arXiv}.

\section*{Data Availability}

Data underlying this article from the DUVET survey will be shared on reasonable request to the PI, Deanne Fisher, at dfisher@swin.edu.au.  Table~\ref{tab:all_values} is available in the online supplementary material. All other data used within this article is archival and available publicly.



\bibliographystyle{mnras}
\bibliography{bibliography} 




\appendix

\section{Baseline Corrections Around \texorpdfstring{H$\beta$}{TEXT}}
\label{appendix:baseline_corr}

\begin{figure*}
    \centering
    \includegraphics[width=\linewidth]{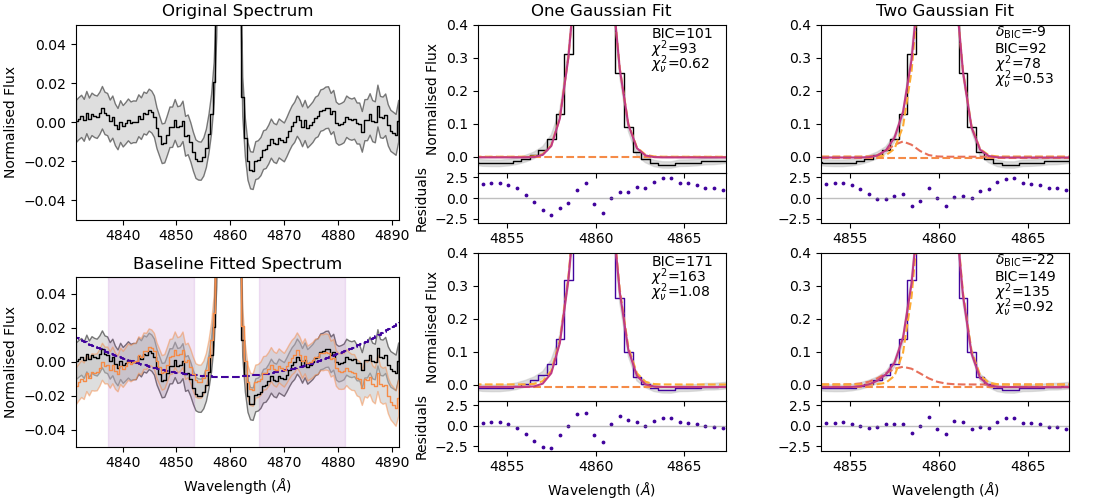}
    \caption{An example of the baseline fitting routine we follow for H$\beta$.  \textit{Top left panel}: original normalised emission line (black line) with variance (shaded grey).  \textit{Bottom left panel}: original emission line (black) is fit with a quadratic baseline (dashed purple) using the points in the purple shaded regions to either side of the H$\beta$ emission line.  The baseline is subtracted, and the resulting emission line is shown in yellow.
    The remaining panels show the fits using a simple one Gaussian + constant model (middle panels); and a two Gaussian + constant model (right panels) for the original and baseline-corrected emission lines (top row and bottom row respectively).  Components of the fit (Gaussians and constant) are shown as yellow and orange dashed lines; the bestfit model is solid magenta.  Residuals from the bestfit model before normalisation are shown in the bottom section of each panel.  Applying the quadratic baseline correction improves the residuals of the double Gaussian fit for H$\beta$.}
    \label{fig:baseline_corr_Hbeta}
\end{figure*}

\begin{figure*}
    \centering
    \includegraphics[width=\linewidth]{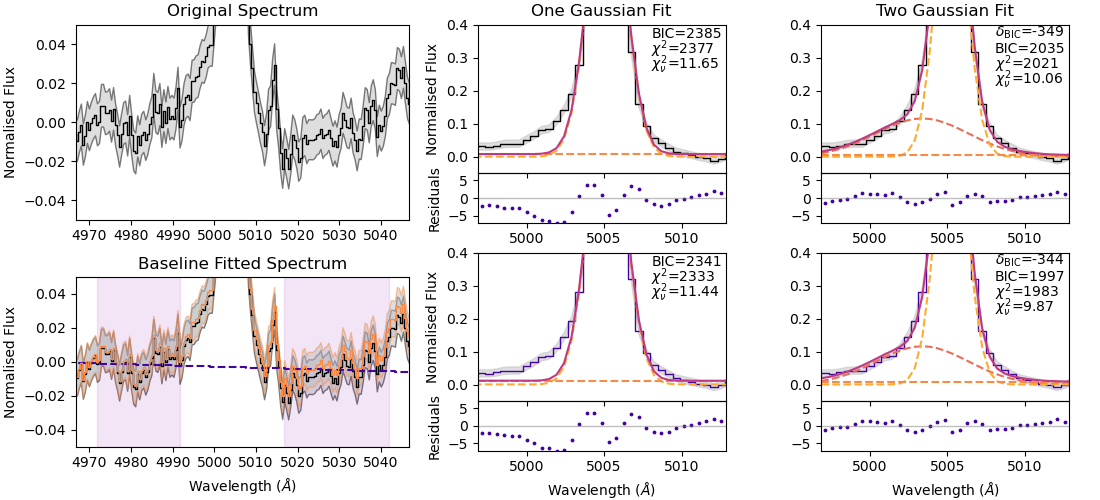}
    \caption{As described for Fig.~\ref{fig:baseline_corr_Hbeta}, using results for [OIII]~$\lambda$5007 from the same spaxel with a linear fit across the baseline in the bottom row.  Applying the linear baseline correction does not make a significant impact on the resulting outflow fit for [OIII]~$\lambda$5007.}
    \label{fig:baseline_corr_OIII}
\end{figure*}

A general continuum subtraction was applied by fitting the stellar continuum in each spaxel with the full spectrum fitting code \texttt{pPXF} \citep{cappellari2017ppxf} and then subtracting this from the spectrum. For more detail, see Section~\ref{subsec:cont_subtract}.  Although we include multiplicative polynomials with high orders (20-25), we identify some spaxels where the wide absorption feature near the H$\beta$ emission line is not fully removed.  We, therefore, apply an extra baseline correction.  We fit a quadratic polynomial to two short bandpass regions on either side of the H$\beta$ emission, carefully chosen for each galaxy to avoid the emission line itself. We additionally fit a linear polynomial to two short bandpass regions on either side of the [OIII]~$\lambda$5007 emission line, carefully chosen to avoid the both the [OIII]~$\lambda$5007 and FeII~$\lambda$5018 emission lines. The resulting baseline fit is then subtracted from the spectrum.  Applying the quadratic baseline correction improves the residuals of the double Gaussian fit for H$\beta$, but does not make a significant difference for [OIII]~$\lambda$5007. An example of this fitting process is given in Fig.~\ref{fig:baseline_corr_Hbeta} for H$\beta$ and in Fig.~\ref{fig:baseline_corr_OIII} for [OIII]~$\lambda$5007.

\section{Individual Galaxy Correlations}
\label{appendix:ind_gals}

\begin{table*}
    \centering
    \small
    \begin{tabular}{lccccccccccccc}
    \hline
        Galaxy Name & \multicolumn{2}{c}{$v_{\rm out, [OIII]}$} & \multicolumn{2}{c}{$v_{\rm out, H\beta}$} & \multicolumn{3}{c}{$\dot{M}_{\rm out}$} & \multicolumn{2}{c}{$\dot{\Sigma}_{\rm out}$} & \multicolumn{3}{c}{$\eta$} \\
         & \multicolumn{2}{c}{(km~s$^{-1}$)} & \multicolumn{2}{c}{(km~s$^{-1}$)} & \multicolumn{3}{c}{($M_\odot$~yr$^{-1}$)} & \multicolumn{2}{c}{($M_\odot$~yr$^{-1}$~kpc$^{-2}$)} & & & \\
         \cmidrule(lr){2-3}\cmidrule(lr){4-5}\cmidrule(lr){6-8}\cmidrule(lr){9-10}\cmidrule(lr){11-13}
         & Median & $\sigma$ & Median & $\sigma$ & Total & Median & $\sigma$ & Median & $\sigma$ & Total & Median & $\sigma$ \\
        \hline
        IRAS~08339+6517 & 286 & 82.6 & 285 & 75.3 & 12.8 & 0.04 & 0.09 & 0.65 & 1.58 & 1.88 & 2.49 & 2.66 \\
        IRAS~15229+0511 & -- & -- & 235 & 58.1 & 0.50 & 0.02 & 0.02 & 0.06 & 0.05 & 0.11 & 0.48 & 0.46 \\
        KISSR~1084 & 325 & 93.1 & 277 & 28.1 & 0.19 & 0.01 & 0.01 & 0.03 & 0.03 & 0.07 & 0.36 & 0.18 \\
        NGC~7316 & 136 & -- & 86.6 & 41.8 & 0.13 & 0.01 & 0.01 & 0.03 & 0.04 & 0.09 & 0.60 & 0.63 \\
        UGC~01385 & 424 & 113 & 294 & 76.3 & 9.91 & 0.08 & 0.38 & 0.76 & 0.13 & 1.74 & 2.37 & 2.57 \\
        UGC~10099 & 240 & 98.9 & 198 & 55.5 & 2.43 & 0.02 & 0.03 & 0.04 & 0.09 & 0.52 & 0.73 & 0.78 \\
        CGCG~453-062 & 502 & 633 & 251 & 232 & 1.64 & 0.03 & 0.03 & 0.15 & 0.13 & 0.11 & 0.56 & 0.74 \\
        UGC~12150 & -- & -- & 305 & 143 & 0.17 & 0.01 & 0.01 & 0.06 & 0.08 & 0.05 & 1.28 & 1.17 \\
        IRAS~20351+2521 & 202 & 187 & 160 & 95.4 & 1.52 & 0.04 & 0.04 & 0.12 & 0.11 & 0.10 & 0.32 & 0.63 \\
        NGC~0695 & 194 & 60.8 & 168 & 120 & 2.04 & 0.03 & 0.03 & 0.09 & 0.10 & 0.06 & 0.44 & 0.56 \\
        \hline 
    \end{tabular}
    \caption{The median and 1$\sigma$ scatter for the maximum outflow velocity $v_{\rm out}$ measured from the [OIII]~$\lambda5007$ and H$\beta$ emission lines; the total, median and 1$\sigma$ scatter for the mass outflow rate $\dot{M}_{\rm out}$; the median and 1$\sigma$ scatter for the mass outflow flux $\dot{\Sigma}_{\rm out}$; and the total, median and 1$\sigma$ scatter for the mass loading factor $\eta$ within $r_{90}$ for 10 face-on galaxies in the DUVET sample.}
    \label{tab:individual_gals}
\end{table*}

In this appendix, we show the correlations between the SFR surface density and the maximum outflow velocity from fits to the H$\beta$ and [OIII]~$\lambda5007$ emission lines, and the mass outflow flux for each individual galaxy in our sample.  In each figure, the panels are organised from left to right and then top to bottom, such that the lowest total galaxy mass is in the top left panel and the highest total galaxy mass in the bottom right panel.  Total galaxy masses for the galaxies are taken from literature values, and are given in Table~\ref{tab:gal_params}.  The median and 1$\sigma$ scatter values for the outflow velocities and mass outflow flux, and the total, median and 1$\sigma$ scatter values for the non-extinction corrected mass outflow rate and mass loading factor within $r_{90}$ for each galaxy are given in Table~\ref{tab:individual_gals}.  All values for spaxels with evidence of outflows in either [OIII]~$\lambda5007$ or H$\beta$ across all 10 galaxies in our sample are given in Table~\ref{tab:all_values}, with both extinction corrected and non-extinction corrected values for the mass outflow rate, mass outflow flux and mass loading factors included.  For the full version of the table, see the online Supplementary Material.

\begin{figure*}
    \centering
    \includegraphics[width=0.9\textwidth]{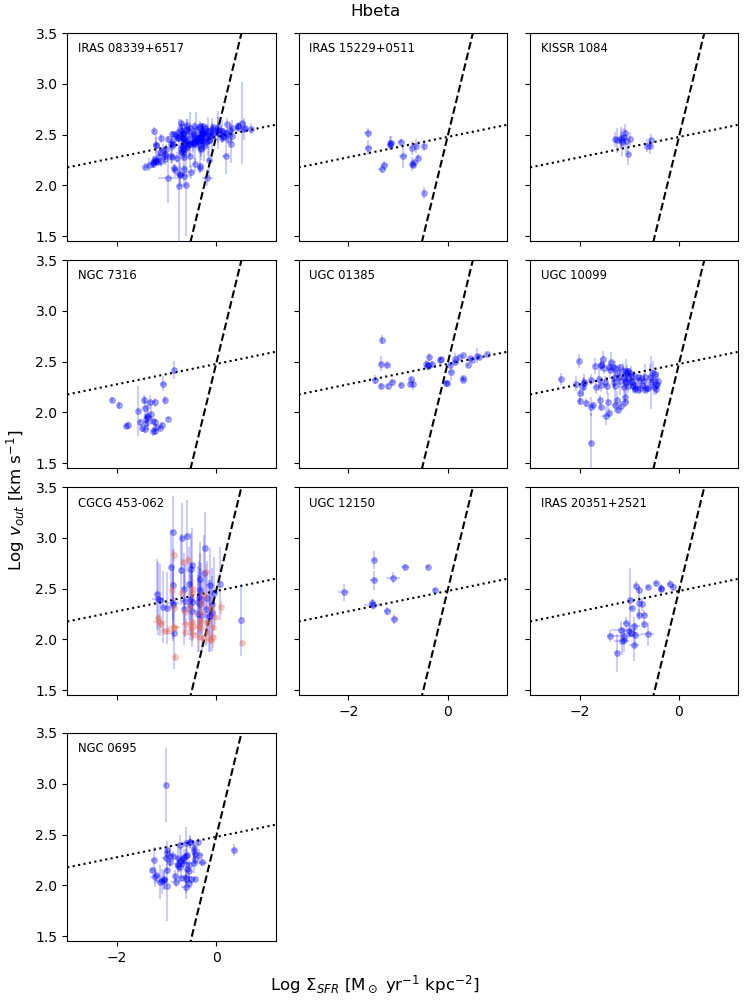}
    \caption{The maximum outflow velocity, $v_{\rm out}$, calculated using the H$\beta$ line plotted against the star formation rate surface density, $\Sigma_{\rm SFR}$ for individual galaxies in our sample, given in order of increasing total stellar mass from left to right, top to bottom.  Orange points show the velocities we observe before reprojection.  The dashed lines represent a model where outflows are primarily driven by momentum injected into the gas from young massive stars \citep[$v_{\rm out}\propto\Sigma_{\rm SFR}^2$;][]{murray2011radiationpressure}.  The dotted lines represent a model where outflows are primarily driven by energy injected from supernovae \citep[$v_{\rm out}\propto\Sigma_{\rm SFR}^{0.1}$;][]{chen2010absorption}.}
    \label{fig:sigsfr_vout_individual_hbeta}
\end{figure*}

\begin{figure*}
    \centering
    \includegraphics[width=0.9\textwidth]{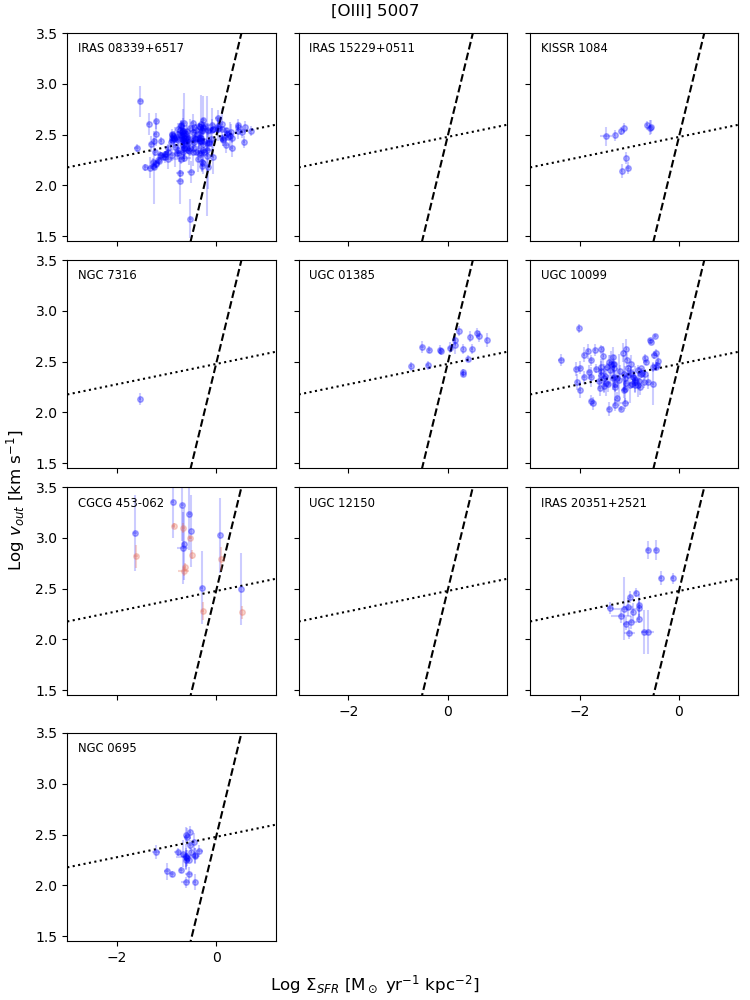}
    \caption{As for Fig.~\ref{fig:sigsfr_vout_individual_hbeta}, but using results from fits to the [OIII]~$\lambda5007$ emission line.}
    \label{fig:sigsfr_vout_individual_OIII}
\end{figure*}

In Fig.~\ref{fig:sigsfr_vout_individual_hbeta} and Fig.~\ref{fig:sigsfr_vout_individual_OIII} we compare the maximum outflow velocity to the SFR surface density for H$\beta$ and [OIII]~$\lambda5007$ respectively for each galaxy in our sample.  All of the galaxies in our sample except for IRAS~08339+6517 are brighter in H$\beta$ than in [OIII]~$\lambda5007$.
Two of our galaxies (UGC~12150 and IRAS~15229+0511
) do not have the signal-to-noise necessary to fit an outflow component in [OIII]~$\lambda5007$.  In addition, we find only one spaxel in NGC~7316 where [OIII]~$\lambda5007$ is fit with an outflow component. 

CGCG~453-062's high inclination angle ($i\approx54^\circ$) required that the velocities be deprojected by dividing by $\cos{i}$.  The originally observed velocities are shown in the CGCG~453-062 panels of Fig.~\ref{fig:sigsfr_vout_individual_hbeta} and Fig.~\ref{fig:sigsfr_vout_individual_OIII} as transparent orange circles.  The deprojected velocities are shown as blue points.
The median H$\beta$ outflow velocity before correction is 132~km~s$^{-1}$, and the inclination correction increases this to 251~km~s$^{-1}$.

In Fig.~\ref{fig:sigsfr_mout_allgals} we compare the ionised gas mass outflow flux to the SFR surface density for each galaxy in our sample.  We show here the results for the outflow flux with no extinction correction applied (see Section~\ref{subsec:mass_flux} for more information on the extinction correction).  We find that the majority of galaxies in our sample follow a trend of increasing outflow mass flux with increasing SFR surface density.  This suggests that the important global change between galaxies is the degree of star formation driving the outflows, but on smaller scales the same local relationship remains.

\begin{figure*}
    \centering
    \includegraphics[width=0.9\textwidth]{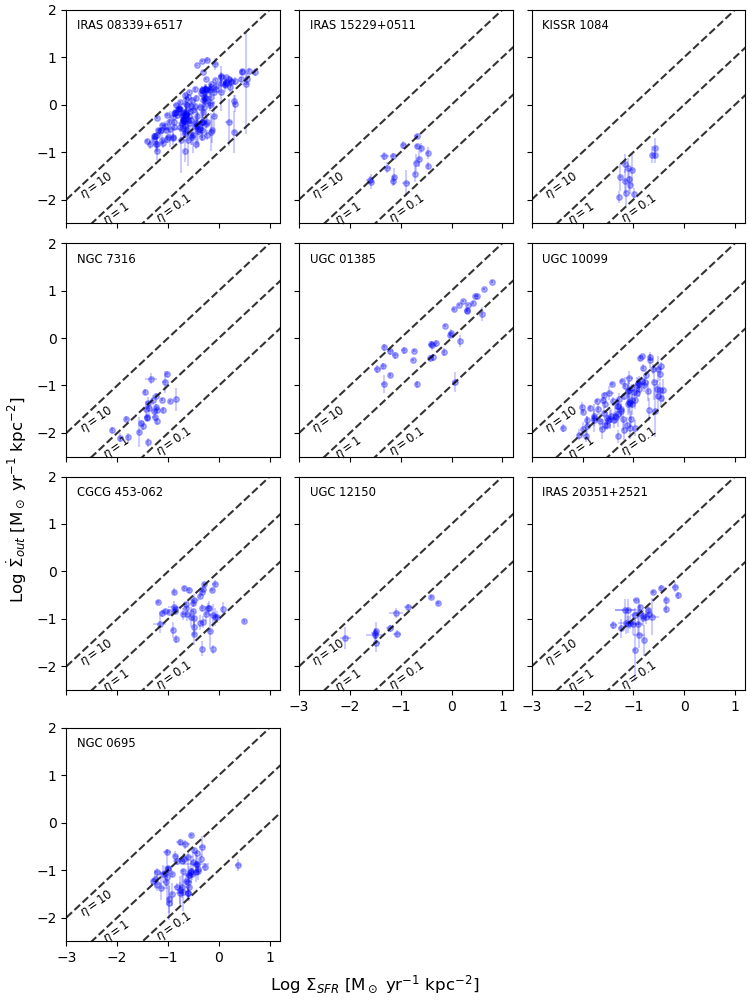}
    \caption{The ionised gas outflow mass flux, $\dot{\Sigma}_{\rm out}$, compared to the SFR surface density, $\Sigma_{\rm SFR}$, for spaxels of sub-kpc resolution pixels in local starbursting disk galaxies with evidence for outflow components.  Each panel gives results for a different galaxy in our sample, given in order of increasing total stellar mass from left to right, top to bottom. Dashed black lines show constant mass loading factors.}
    \label{fig:sigsfr_mout_individual}
\end{figure*}

\begin{landscape}
\begin{table}
    \centering
    \caption{Values for all spaxels containing evidence of outflows in either [OIII]~$\lambda$5007 or H$\beta$.  The galaxy name, star formation rate surface density, stellar mass surface density, outflow velocity in H$\beta$ and [OIII]~$\lambda$5007, outflow mass rate, outflow mass flux, and mass loading factor.  We also include the values for the outflow mass rate, outflow mass flux and mass loading factor when assuming the same extinction in the outflow gas as is measured for the disk gas.  For a full table, see the online Supplementary Material.}
    \label{tab:all_values}
    \small
    \begin{tabular}{lllllllllll}
    \toprule
    Galaxy Name & $\Sigma_{\rm SFR}$ & $\Sigma_\ast$ & $v_{\rm out, H\beta}$ & $v_{\rm out, OIII}$ & $\dot{M}_{\rm out}$ & $\dot{M}_{\rm out, ext corr}$ & $\dot{\Sigma}_{\rm out}$ & $\dot{\Sigma}_{\rm out, ext corr}$ & $\eta$ & $\eta_{\rm ext corr}$ \\
     & (M$_\odot$~yr$^{-1}$~kpc$^2$) & (M$_\odot$~pc$^2$) & (km~s$^{-1}$) & (km~s$^{-1}$) & (M$_\odot$~yr$^{-1}$) & (M$_\odot$~yr$^{-1}$) & (M$_\odot$~yr$^{-1}$~kpc$^2$) & (M$_\odot$~yr$^{-1}$~kpc$^2$) &  &  \\
    \midrule
    IRAS 08339+6517 & 0.14$\pm$0.00 & 1.00e4$\pm$1.40e4 & -- & 2.10e2$\pm$25 & -- & -- & -- & -- & -- & -- \\
     & 0.14$\pm$0.01 & 9.90e3$\pm$1.30e4 & -- & 2.90e2$\pm$50 & -- & -- & -- & -- & -- & -- \\
     & 0.17$\pm$0.01 & 1.20e4$\pm$1.70e4 & -- & 2.20e2$\pm$35 & -- & -- & -- & -- & -- & -- \\
     & 0.21$\pm$0.01 & 1.40e4$\pm$1.90e4 & 2.50e2$\pm$52 & 2.10e2$\pm$34 & 0.01$\pm$0.00 & 0.01$\pm$0.00 & 0.11$\pm$0.06 & 0.11$\pm$0.06 & 0.50$\pm$0.28 & 0.53$\pm$0.30 \\
     & 0.19$\pm$0.00 & 1.50e4$\pm$2.10e4 & -- & 1.70e2$\pm$28 & -- & -- & -- & -- & -- & -- \\
     & 0.30$\pm$0.02 & 1.30e4$\pm$1.60e4 & 2.00e2$\pm$52 & -- & 0.01$\pm$0.00 & 0.03$\pm$0.01 & 0.21$\pm$0.08 & 0.43$\pm$0.17 & 0.69$\pm$0.27 & 1.40$\pm$0.57 \\
     & 0.19$\pm$0.02 & 1.60e4$\pm$2.20e4 & 2.10e2$\pm$33 & 1.80e2$\pm$22 & 0.03$\pm$0.01 & 0.03$\pm$0.01 & 0.48$\pm$0.13 & 0.50$\pm$0.14 & 2.50$\pm$0.72 & 2.60$\pm$0.75 \\
     & 0.43$\pm$0.01 & 1.70e4$\pm$2.40e4 & -- & 2.40e2$\pm$86 & -- & -- & -- & -- & -- & -- \\
     & 0.29$\pm$0.02 & 1.50e4$\pm$1.90e4 & 2.50e2$\pm$74 & 2.20e2$\pm$68 & 0.01$\pm$0.01 & 0.02$\pm$0.01 & 0.23$\pm$0.11 & 0.36$\pm$0.17 & 0.80$\pm$0.39 & 1.30$\pm$0.60 \\
     & 0.22$\pm$0.02 & 1.90e4$\pm$2.60e4 & 2.20e2$\pm$37 & 2.30e2$\pm$58 & 0.04$\pm$0.01 & 0.04$\pm$0.01 & 0.63$\pm$0.19 & 0.66$\pm$0.20 & 2.90$\pm$0.93 & 3$\pm$0.96 \\
     ... & ... & ... & ... & ... & ... & ... & ... & ... & ... & ... \\
    \bottomrule
    \end{tabular} 
\end{table}
\end{landscape}


\bsp	
\label{lastpage}
\end{document}